\documentclass[12pt]{article}
\usepackage{amssymb,amsmath,epsfig}
\allowdisplaybreaks

\begin{document}

\title{\bf Impact of Charge on Complexity Analysis and Isotropic Decoupled Solutions in $f(\mathbb{R},\mathbb{T})$ Gravity}
\author{M. Sharif$^1$ \thanks{msharif.math@pu.edu.pk}~ and Tayyab Naseer$^{1,2}$
\thanks{tayyabnaseer48@yahoo.com}\\
$^1$ Department of Mathematics and Statistics, The University of Lahore,\\
1-KM Defence Road Lahore-54000, Pakistan.\\
$^2$ Department of Mathematics, University of the Punjab,\\
Quaid-i-Azam Campus, Lahore-54590, Pakistan.}

\date{}
\maketitle

\begin{abstract}
In this paper, we formulate two exact charged solutions to the field
equations by extending the domain of existing anisotropic models
with the help of minimal gravitational decoupling in
$f(\mathbb{R},\mathbb{T})$ theory. For this, the anisotropic fluid
distribution is considered as a seed source that is extended through
the inclusion of a new gravitational source. The influence of the
later matter configuration is controlled by the decoupling
parameter. We formulate the field equations corresponding to the
total matter source that are then decoupled into two distinct sets
by implementing a transformation only on the radial metric
coefficient. Both of these under-determined sets correspond to their
parent sources. Some well-behaved forms of the metric potentials are
taken into account to deal with the first set of equations. On the
other hand, we solve the second set corresponding to an additional
source by taking different constraints on the matter sector. We then
consider the radius and mass of a compact star $4U~1820-30$ to
analyze the physical feasibility of the resulting solutions for a
particular modified model. It is concluded that our resulting
solutions show stable behavior for certain values of the decoupling
parameter and charge.
\end{abstract}

\section{Introduction}

Cosmological breakthroughs show that astrophysical structures are
not scattered randomly in our cosmos but are systematically
distributed. The analysis of such an integrated paradigm as well as
different physical characteristics of compact self-gravitating
objects enable astronomers to expose the accelerated expansion of
the universe. Multiple modifications to general relativity
($\mathbb{GR}$) have been suggested to explain such a cosmic
expansion in the recent couple of years. The insertion of the
generic function of the Ricci scalar in place of $\mathbb{R}$ in the
Einstein-Hilbert action produces the first and straightforward
extension of $\mathbb{GR}$, termed the $f(\mathbb{R})$ theory.
Several researchers made an initial attempt to explain different
evolutionary eras of the universe by adopting modified
$f(\mathbb{R})$ models \cite{2,2a}. The stability of this theory has
also been addressed through different techniques \cite{9,9g}.

Bertolami et al. \cite{10} initially merged the geometric terms with
the matter Lagrangian density in $f(\mathbb{R})$ framework to
produce the effects of fluid-geometry coupling on the test particles
of massive bodies. Several researchers studied such coupling and
concluded that it would be advantageous to discuss the cosmic
accelerated expansion in this framework. Harko et al. \cite{20}
generalized this interaction by introducing a generic function of
$\mathbb{R}$ and trace of the energy-momentum tensor
($\mathbb{EMT}$) $\mathbb{T}$ in the action, referred the
$f(\mathbb{R},\mathbb{T})$ theory. The coupling between geometry and
matter in this framework provides non-vanishing divergence of the
$\mathbb{EMT}$. An extra force thus appears in the gravitational
field that changes the geodesic motion of the particles under
consideration. Different gravitational aspects have been comprised
by this theory due to the entanglement of the entity $\mathbb{T}$.
Houndjo \cite{22a} considered a minimal model of this gravity and
explained the transformation of the matter-dominated phase into
late-time acceleration era. Das et al. \cite{22b} adopted the
$\mathbb{R}+2\chi_3\mathbb{T}$ model to explore the structure of
gravastar comprising three different layers which are expressed by
their respective equations of state. Different matter distributions
have been discussed for such modified models from which various
acceptable solutions representing compact interiors are obtained
\cite{25aa}-\cite{25c}.

It is worth noting that the problem of the cosmological constant
($\Lambda$) can considerably be resolved through a particular choice
such as $\mathbb{R}+2\chi_3\mathbb{T}$. It has been shown that the
constant $\Lambda$ can be expressed in proportional relation with
the Hubble parameter
($\mathbb{H}=\frac{\dot{\mathbf{a}}}{\mathbf{a}}$, where
$\mathbf{a}$ denotes the scale factor of the cosmos), i.e.,
$\Lambda_{eff} \propto \mathbb{H}^2$ \cite{1i}. The second term in
the above model is a valid one that could generate several new
insights and helps to understand the astronomical structures in a
proper way \cite{40a,40aa}. The model parameter has been restricted
as $\chi_3 > -3.0\times10^{-4}$ while studying massive white dwarfs
\cite{aa}. In addition, the background evolution provided a limit on
this parameter as $-0.1 < \chi_3 < 1.5$ \cite{bb}. Also, the lower
bound of $\chi_3$ was obtained as $-1.9\times10^{-8}$ from the dark
energy density parameter \cite{cc}. Ashmita et al. \cite{37n}
adopted three different inflation potentials in this modified
theory, and derived the corresponding field equations and potential
slow-roll parameters. They concluded that their results are
compatible with the observational data only for
$-0.37<\chi_3<1.483$. Kaur et al. \cite{38} assumed this model to
discuss the interior of charged anisotropic Tolman-Kuchowicz
solution and found it stable as well as viable.

The nature of compact stellar bodies in our universe can be
understood by formulating the solutions (exact or numerical) of
their respective field equations. The highly non-linear nature of
such equations prompted several astrophysicists to find different
techniques so that these equations can be solved easily which
further helps to check the physical relevancy of the respective
model. One of the multiple approaches discussed in the literature is
the gravitational decoupling that helps to obtain the feasible
solution of the interior fluid distribution possessing multiple
factors like pressure anisotropy, heat dissipation and shear stress.
This approach solves the field equations corresponding to more than
one matter configuration by transforming the metric components to a
new frame of reference which makes much easier to solve them.
Gravitational decoupling offers two schemes, one of them is the
minimal geometric deformation (MGD). Ovalle \cite{29} recently
developed this approach and found the analytic solutions
representing stellar objects in the braneworld (BW). Later, Ovalle
and Linares \cite{30} formulated an exact solution to the field
equations for an isotropic sphere by using the MGD technique along
with the Tolman-IV metric potentials in the BW scenario. Casadio et
al. \cite{31} developed a consistent extension of the above
technique and proposed a new solution in the context of BW that
helps to discuss the geometry of stars.

Ovalle et al. \cite{33} employed the gravitational decoupling
through MGD approach and extended an isotropic solution to the
anisotropic domain. They also analyzed the feasibility of resulting
solutions through graphical interpretation. These isotropic
solutions were then extended to the case of charged Krori-Barua
anisotropic sphere in which the influence of an electromagnetic
field on the stability of resulting models has been checked
\cite{34}. Gabbanelli et al. \cite{36} developed the anisotropic
Duragpal-Fuloria models representing physically relevant celestial
objects. The same analysis has also been done for multiple
extensions of the isotropic Heintzmann as well as Tolman VII
spacetimes \cite{36a,37a}. The MGD approach was also used to extend
several isotropic spacetimes to an acceptable anisotropic domain in
Brans-Dicke theory \cite{37c}. Multiple compact models have been
developed in a non-minimal gravitational theory through the
decoupling technique \cite{37f}. The influence of the respective
parameter was also checked on the corresponding physical
determinants.

The analysis of self-gravitating celestial systems has been observed
to be significantly affected by the presence of an electric charge.
A strongly attractive nature of gravity can be reduced or
counterbalanced through some important ingredients, one of them is
the electromagnetic force. For this reason, a compact body needs a
sufficient amount of the charge in order to resist the gravitational
attraction and preserve its stable state. This has been confirmed by
Bekenstein \cite{37aba} while studying a charged dynamical sphere.
He concluded that the nature of the electromagnetic force is
opposite to that of gravity, helping the considered geometry to be
stable. Esculpi and Aloma \cite{37abb} have extended this work to
explore the behavior of anisotropic matter distribution in a compact
interior. They observed that the spacetime experiences a repulsive
force due to the presence of both charge and positive anisotropy.
Some anisotropic charged compact stars have been studied by adopting
a particular form of the interior charge in terms of the spherical
radius and acceptable results were obtained \cite{37abc,37abd}.
Pretel et al. \cite{37abe} explored the interior distribution of
charged quark stars through a linear equation of state in the
context of $\mathbb{R}+2\chi_3\mathbb{T}$ gravity. They observed
that the parameters measuring the impact of charge and modified
theory considerably affect the total mass and radius of the star.
Several physically relevant charged compact stars have been modeled
by us in a non-minimally coupled framework \cite{37abf}.

Herrera \cite{37g} suggested a very recent definition of the
complexity for static self-gravitation spherical systems. He used
Bel's idea for the orthogonal decomposition of the curvature tensor
corresponding to the anisotropic matter configuration and found
several scalars. It was then observed that there is only one scalar
that involves the inhomogeneous density and anisotropy, thus
referred as the complexity factor. This definition becomes widely
accepted for the last couple of years in the scientific community.
Herrera with his collaborators also extended this definition for
dynamical matter source \cite{37h}. Such extensions have been
developed in non-minimally coupled theory for static/non-static
spherical and cylindrical interiors \cite{37i,37ia}. Different
constraints have been used in the literature to make the system
solvable. Several researchers \cite{37j} used the complexity-free
condition as a constraint and merged it with the decoupling strategy
to obtain physically relevant stellar models. Casadio et al.
\cite{37k} considered anisotropic compact model and isotropized it
by varying the decoupling parameter involving in the MGD technique.
Maurya et al. \cite{37l} adopted the no complexity constraint and
checked the effect of the decoupling parameter on the developed
embedding class-one model. Sharif and Majid \cite{37m} also
formulated such models in Brans-Dicke theory and found them stable
for certain parametric values.

In this article, we investigate how two different solutions
representing compact models are affected by the charge as well as
modified corrections of $f(\mathbb{R},\mathbb{T})$ gravity.
Following lines depict how this paper is organized. We introduce
some basic foundations of the modified theory and formulate the
corresponding field equations in the presence of charge as well as
an extra matter source in section \textbf{2}. Section \textbf{3}
separates these equations into two sets through the MGD scheme. Two
different constraints depending on the matter sector are adopted in
sections \textbf{4} and \textbf{5}, leading to the resulting
solutions. Section \textbf{6} analyzes the corresponding matter
determinants and other physical characteristics to check whether the
modified models are physically relevant or not. Lastly, section
\textbf{7} sums up our results.

\section{$f(\mathbb{R},\mathbb{T})$ Gravity}

The presence of an additional source in the gravitational field
modifies the Einstein-Hilbert action (with $\kappa=8\pi$) in the
following form \cite{20}
\begin{equation}\label{g1}
\mathrm{S}=\int\sqrt{-g}\left[\frac{f(\mathbb{R},\mathbb{T})}{16\pi}+\mathbb{L}_{m}+\mathbb{L}_{\mathbb{E}}
+\gamma\mathbb{L}_{\mathfrak{C}}\right]d^{4}x,
\end{equation}
where $\mathbb{L}_{\mathfrak{C}}$ being the Lagrangian density of an
extra source gravitationally coupled to the original field,
$\mathbb{L}_{m}$ and $\mathbb{L}_{\mathbb{E}}$ are that of the
matter distribution and an electromagnetic field, respectively.
Also, the determinant of the metric tensor $g_{\beta\alpha}$ is
symbolized by $g$. Taking the action \eqref{g1} and applying the
principle of least-action produces the field equations as
\begin{equation}\label{g2}
\mathbb{G}_{\beta\alpha}=8\pi
\mathbb{T}_{\beta\alpha}^{(\mathrm{total})},
\end{equation}
where $\mathbb{G}_{\beta\alpha}$ being an Einstein tensor that
describes the geometry of the spacetime whereas the interior fluid
distribution is characterized by the right hand side of the above
equation. We further classify it for the current scenario as
\begin{equation}\label{g3}
\mathbb{T}_{\beta\alpha}^{(\mathrm{total})}=\mathbb{T}_{\beta\alpha}^{(\mathrm{ef})}+\gamma
\mathfrak{C}_{\beta\alpha}=\frac{1}{f_{\mathbb{R}}}\big(\mathbb{T}_{\beta\alpha}+\mathbb{E}_{\beta\alpha}\big)
+\mathbb{T}_{\beta\alpha}^{(\mathrm{D})}+\gamma\mathfrak{C}_{\beta\alpha},
\end{equation}
where $f_{\mathbb{R}}=\frac{\partial
f(\mathbb{R},\mathbb{T})}{\partial \mathbb{R}}$. Moreover, $\gamma$
controls the effect of an additional matter sector
($\mathfrak{C}_{\beta\alpha}$) on the seed interior configuration of
celestial systems, and we call it the decoupling parameter. The
modification in the Einstein-Hilbert action results in the effective
$\mathbb{EMT}$, denoted by
$\mathbb{T}_{\beta\alpha}^{(\mathrm{ef})}$ containing the usual
matter part of $\mathbb{GR}$ \big(in which
$\mathbb{T}_{\beta\alpha}$ and $\mathbb{E}_{\beta\alpha}$ being the
usual and electromagnetic $\mathbb{EMT}s$, respectively\big) and
correction terms $\mathbb{T}_{\beta\alpha}^{(\mathrm{D})}$ of the
extended theory. The later term has the form
\begin{eqnarray}
\nonumber \mathbb{T}_{\beta\alpha}^{(\mathrm{D})}&=&\frac{1}{8\pi
f_{\mathbb{R}}}\bigg[\mathbb{T}_{\beta\alpha}f_{\mathbb{T}}+\bigg\{\frac{1}{2}\big(f-\mathbb{R}f_{\mathbb{R}}\big)
-\mathbb{L}_{m}f_{\mathbb{T}}\bigg\}g_{\beta\alpha}\\\label{g4}
&-&(g_{\beta\alpha}\Box-\nabla_{\beta}\nabla_{\alpha})f_{\mathbb{R}}+2g^{\rho\zeta}f_{\mathbb{T}}\frac{\partial^2
\mathbb{L}_{m}}{\partial g^{\beta\alpha}\partial
g^{\rho\zeta}}\bigg],
\end{eqnarray}
where $f_{\mathbb{T}}=\frac{\partial
f(\mathbb{R},\mathbb{T})}{\partial \mathbb{T}}$. Also, the term
$\nabla_\beta$ being the covariant derivative and $\Box\equiv
\frac{1}{\sqrt{-g}}\partial_\beta\big(\sqrt{-g}g^{\beta\alpha}\partial_{\alpha}\big)$
is the D'Alembertian operator.

We shall discuss the complexity of a self-gravitating system,
therefore, the nature of seed fluid distribution must be
anisotropic. The following $\mathbb{EMT}$ represents such kind of
fluids as
\begin{equation}\label{g5}
\mathbb{T}_{\beta\alpha}=\mu\mathcal{K}_{\beta}\mathcal{K}_{\alpha}+P_\bot\big(\mathcal{K}_{\beta}\mathcal{K}_{\alpha}
+g_{\beta\alpha}-\mathcal{W}_\beta\mathcal{W}_\alpha\big)+P_r\mathcal{W}_\beta\mathcal{W}_\alpha,
\end{equation}
where the triplet ($P_r,P_\bot,\mu$) symbolizes the
radial/tangential pressure and density, respectively. Further,
$\mathcal{K}_{\beta}$ being the four-velocity and
$\mathcal{W}_{\beta}$ is the four-vector satisfying the following
relations
\begin{equation}\nonumber
\mathcal{W}^\beta \mathcal{K}_{\beta}=0, \quad \mathcal{W}^\beta
\mathcal{W}_{\beta}=1, \quad \mathcal{K}^\beta
\mathcal{K}_{\beta}=-1.
\end{equation}
Further, the expression for $\mathbb{E}_{\beta\alpha}$ is provided
by
\begin{equation}\nonumber
\mathbb{E}_{\beta\alpha}=-\frac{1}{4\pi}\left[\frac{1}{4}g_{\beta\alpha}\mathbb{F}^{\zeta\eta}\mathbb{F}_{\zeta\eta}
-\mathbb{F}^{\zeta}_{\beta}\mathbb{F}_{\alpha\zeta}\right].
\end{equation}
Here, the Maxwell field tensor is defined as
$\mathbb{F}_{\zeta\eta}=\Phi_{\eta;\zeta}-\Phi_{\zeta;\eta}$ with
$\Phi_{\zeta}=\Phi(r)\delta^{0}_{\zeta}$ being the four-potential. A
concise form of the Maxwell equations is presented in the following
\begin{equation}\label{g5b}
\mathbb{F}^{\zeta\eta}_{;\eta}=4\pi \jmath^{\zeta}, \quad
\mathbb{F}_{[\zeta\eta;\alpha]}=0,
\end{equation}
where $\jmath^{\zeta}=\Omega\mathcal{K}^{\zeta}$ and $\Omega$ being
the current and charge densities, respectively.

The extended field equations \eqref{g2} have the following trace
given by
\begin{align}\nonumber
&\big(\mathbb{R}+3\nabla^{\beta}\nabla_{\beta}\big)f_\mathbb{R}-\mathbb{T}+\big(\mathbb{T}+4\mathbb{L}_m\big)f_\mathbb{T}-2f
-2f_\mathbb{T}g^{\rho\zeta}g^{\beta\alpha}\frac{\partial^2\mathbb{L}_m}{\partial
g^{\rho\zeta}\partial g^{\beta\alpha}}=0.
\end{align}
By taking the vacuum case into account, we can retrieve the
$f(\mathbb{R},\mathbb{T})$ equations of motion and their solutions
to $f(\mathbb{R})$ gravitational theory. The covariant divergence of
effective $\mathbb{EMT}$ is zero, however, that of usual matter
distribution is non-null due to the inclusion of the
curvature-matter coupling in the gravitational action. Consequently,
an additional force comes into existence that can be seen from the
following expression
\begin{align}\nonumber
\nabla^\beta\mathbb{T}_{\beta\alpha}&=\frac{f_\mathbb{T}}{8\pi-f_\mathbb{T}}\bigg[(\mathbb{T}_{\beta\alpha}
+\Theta_{\beta\alpha})\nabla^\beta\ln{f_\mathbb{T}}+\nabla^\beta\Theta_{\beta\alpha}\\\label{g11}
&-\frac{8\pi\gamma}{f_\mathbb{T}}\nabla^\beta\mathfrak{C}_{\beta\alpha}
-\frac{1}{2}g_{\rho\zeta}\nabla_\alpha\mathbb{T}^{\rho\zeta}-\frac{8\pi}{f_\mathbb{T}}\nabla^\beta\mathbb{E}_{\beta\alpha}\bigg],
\end{align}
where
$\Theta_{\beta\alpha}=g_{\beta\alpha}\mathbb{L}_m-2\mathbb{T}_{\beta\alpha}-2g^{\rho\zeta}\frac{\partial^2
\mathbb{L}_{m}}{\partial g^{\beta\alpha}\partial g^{\rho\zeta}}$.
Several choices of the matter Lagrangian have been analyzed in the
literature for isotropic/anisotropic fluids including the energy
density, pressure, torsion, etc. Since we consider the anisotropic
matter distribution, $\mathbb{L}_{m}=P=\frac{2P_\bot+P_r}{3}$ is the
most suitable choice in this case \cite{20}.

A self-gravitating geometry can be distinguished into an interior
and exterior regions by the hypersurface $\Sigma$. We define the
interior region of a static spherical spacetime by the following
line element as
\begin{equation}\label{g6}
\mathrm{ds}^2=-e^{\chi_1}dt^2+e^{\chi_2}dr^2+r^2d\vartheta^2+r^2\sin^2\vartheta
d\varphi^2,
\end{equation}
where $\chi_1=\chi_1(r)$ and $\chi_2=\chi_2(r)$. The two quantities
$\mathcal{W}_\beta$ and $\mathcal{K}_\beta$ in Eq.\eqref{g5} become
in terms of the above metric as
\begin{equation}\nonumber
\mathcal{W}_\beta=(0,e^{\frac{\chi_2}{2}},0,0), \quad
\mathcal{K}_\beta=(-e^{\frac{\chi_1}{2}},0,0,0).
\end{equation}
Further, Eq.\eqref{g5b} (left) now leads to the following
differential equation
\begin{equation}
\Phi''-\frac{1}{2r}\big\{r(\chi_1'+\chi_2')-4\big\}\Phi'=4\pi\Omega
e^{\frac{\chi_1}{2}+\chi_2},
\end{equation}
whose integration yields
\begin{equation}
\Phi'=\frac{\mathrm{s}(r)}{r^2}e^{\frac{\chi_1+\chi_2}{2}},
\end{equation}
where $'=\frac{\partial}{\partial r}$ and $\mathrm{s}(r)=\int_0^r
\Omega e^{\frac{\chi_2}{2}}\hat{r}^2d\hat{r}$ being the total
interior charge.

We shall present a graphical analysis of the resulting solutions
(will be developed in the later sections), thus it is necessary to
adopt a specific model that can help us to express our results
meaningfully. Several matter-geometry coupled
$f(\mathbb{R},\mathbb{T})$ models have been studied in the
literature. Since we aim to find our solutions through the
gravitational decoupling strategy, a very useful linear model in
this regard is assumed in the following
\begin{equation}\label{g61}
f(\mathbb{R},\mathbb{T})=f_1(\mathbb{R})+
f_2(\mathbb{T})=\mathbb{R}+2\chi_3\mathbb{T},
\end{equation}
where $\mathbb{T}=P_r+2P_\bot-\mu$ and $\chi_3$ being a real-valued
coupling constant. We have employed this model to construct two
different anisotropic solutions and observed our results consistent
with $\mathbb{GR}$ \cite{39}.

The line element \eqref{g6} and modified model \eqref{g61} produce
the non-zero components of the field equations \eqref{g2} as
\begin{align}\label{g8}
8\pi\left(\mu-\gamma\mathfrak{C}_{0}^{0}\right)+\frac{\mathrm{s}^2}{r^4}
&=e^{-\chi_2}\bigg(\frac{\chi_2'}{r}-\frac{1}{r^2}\bigg)+\frac{1}{r^2}
-\chi_3\bigg(3\mu-\frac{P_r}{3}-\frac{2P_\bot}{3}\bigg),\\\label{g9}
8\pi\left(P_r+\gamma\mathfrak{C}_{1}^{1}\right)-\frac{\mathrm{s}^2}{r^4}
&=e^{-\chi_2}\bigg(\frac{1}{r^2}+\frac{\chi_1'}{r}\bigg)-\frac{1}{r^2}
+\chi_3\bigg(\mu-\frac{7P_r}{3}-\frac{2P_\bot}{3}\bigg),\\\nonumber
8\pi\left(P_\bot+\gamma\mathfrak{C}_{2}^{2}\right)+\frac{\mathrm{s}^2}{r^4}
&=\frac{e^{-\chi_2}}{4}\bigg(\chi_1'^2-\chi_2'\chi_1'+2\chi_1''-\frac{2\chi_2'}{r}+\frac{2\chi_1'}{r}\bigg)\\\label{g10}
&+\chi_3\bigg(\mu-\frac{P_r}{3}-\frac{8P_\bot}{3}\bigg),
\end{align}
where the entities along by $\chi_3$ on the right side of above
equations appear due to $f(\mathbb{R},\mathbb{T})$ corrections.
Moreover, Eq.\eqref{g11} together with \eqref{g61} takes the form
\begin{align}\nonumber
&\frac{dP_r}{dr}+\frac{\chi_1'}{2}\left(\mu+P_r\right)+\frac{\gamma\chi_1'}{2}
\left(\mathfrak{C}_{1}^{1}-\mathfrak{C}_{0}^{0}\right)+\frac{2}{r}\left(P_r-P_\bot\right)\\\label{g12}
&+\gamma\frac{d\mathfrak{C}_{1}^{1}}{dr}+\frac{2\gamma}{r}\left(\mathfrak{C}_{1}^{1}
-\mathfrak{C}_{2}^{2}\right)=\frac{1}{\chi_3-4\pi}\bigg\{\chi_3\big(\mu'-P'\big)-\frac{\mathrm{ss}'}{4\pi{r}^4}\bigg\}.
\end{align}
We can call Eq.\eqref{g12} as the generalized
Tolman-Opphenheimer-Volkoff equation that contains different forces
such as Newtonian, hydrodynamical, the gravitational and extra force
due to modified gravity. This equation further helps to study
fluctuations in a self-gravitating structure due to different
factors.

\section{Gravitational Decoupling}

Since the inclusion of a new gravitational source increases unknowns
in the field equations, i.e.,
$(\chi_1,\chi_2,\mathrm{s},\mu,P_\bot,P_r,\mathfrak{C}_{0}^{0},\mathfrak{C}_{1}^{1},\mathfrak{C}_{2}^{2})$,
an analytical solution to this highly non-linear set of equations is
not possible unless we employ different constraints. On that note, a
systematic scheme (termed as the gravitational decoupling \cite{33})
is employed so that the modified field equations can be made
solvable. An interesting point is that this strategy transforms the
metric components and leads the field equations to a new framework
where they can easily be solved. The following solution to
Eqs.\eqref{g8}-\eqref{g10} is now considered to execute this
technique given by
\begin{equation}\label{g15}
ds^2=-e^{\chi_4(r)}dt^2+\frac{1}{\chi_5(r)}dr^2+r^2d\vartheta^2+r^2\sin^2\vartheta
d\varphi^2.
\end{equation}
The linear transformations of the metric components are presented by
\begin{equation}\label{g16}
\chi_4\rightarrow\chi_1=\chi_4+\gamma\mathrm{t}, \quad
\chi_5\rightarrow e^{-\chi_2}=\chi_5+\gamma\mathrm{f},
\end{equation}
where $\mathrm{t}$ being the temporal and $\mathrm{f}$ is the radial
deformation function. We strive to deform only the $g_{rr}$
component, thus MGD scheme is chosen to be applied that preserves
the temporal metric function, i.e., $\mathrm{f} \rightarrow
\bar{\mathcal{F}},~\mathrm{t} \rightarrow 0$. Eq.\eqref{g16} then
switches into
\begin{equation}\label{g17}
\chi_4\rightarrow\chi_1=\chi_4, \quad \chi_5\rightarrow
e^{-\chi_2}=\chi_5+\gamma\bar{\mathcal{F}},
\end{equation}
where $\bar{\mathcal{F}}=\bar{\mathcal{F}}(r)$. It must be noted
that such kind of linear mappings does not disturb the symmetry of a
sphere. The implementation of the transformation \eqref{g17} on the
system \eqref{g8}-\eqref{g10} leads to two distinct sets. The first
of them (corresponding to $\gamma=0$) represents the original
(charged anisotropic) matter source given by
\begin{align}\label{g18}
8\pi\mu+\chi_3\bigg(3\mu-\frac{P_r}{3}-\frac{2P_\bot}{3}\bigg)+\frac{\mathrm{s}^2}{r^4}
&=e^{-\chi_2}\bigg(\frac{\chi_2'}{r}-\frac{1}{r^2}\bigg)+\frac{1}{r^2},\\\label{g19}
8\pi{P}_r-\chi_3\bigg(\mu-\frac{7P_r}{3}-\frac{2P_\bot}{3}\bigg)-\frac{\mathrm{s}^2}{r^4}
&=e^{-\chi_2}\bigg(\frac{1}{r^2}+\frac{\chi_1'}{r}\bigg)-\frac{1}{r^2},\\\label{g20}
8\pi{P}_\bot-\chi_3\bigg(\mu-\frac{P_r}{3}-\frac{8P_\bot}{3}\bigg)+\frac{\mathrm{s}^2}{r^4}
&=\frac{e^{-\chi_2}}{4}\bigg(\chi_1'^2-\chi_2'\chi_1'+2\chi_1''-\frac{2\chi_2'}{r}+\frac{2\chi_1'}{r}\bigg).
\end{align}

The explicit expressions for the triplet ($\mu,P_r,P_\bot$) can be
obtained by solving Eqs.\eqref{g18}-\eqref{g20} simultaneously as
\begin{align}\nonumber
\mu&=\frac{e^{-\chi_2}}{48r^4(\chi_3+2\pi)(\chi_3+4\pi)}\big[r^4
\chi_3  \chi_1 '^2+r^3 \chi_3  \chi_1 ' \big(4-r \chi_2 '\big)+2
\big\{r^4 \chi_3  \chi_1 ''\\\nonumber &+8 r^3 (\chi_3 +3 \pi )
\chi_2 '+24 \pi r^2 e^{\chi_2 }+8 r^2 \chi_3  e^{\chi_2 }-8 r^2
\chi_3 -10 \mathrm{s}^2 \chi_3 e^{\chi_2 }\\\label{g18a} &-24 \pi
r^2-24 \pi \mathrm{s}^2 e^{\chi_2 }\big\}\big],\\\nonumber
P_r&=\frac{e^{-\chi_2}}{48r^4(\chi_3+2\pi)(\chi_3+4\pi)}\big[r^3
\chi_1 ' \big(r \chi_3  \chi_2 '+20 \chi_3 +48 \pi \big)-2 r^4
\chi_3  \chi_1 ''\\\nonumber &-r^4\chi_3\chi_1 '^2+8 r^3 \chi_3
\chi_2 '-48 \pi  r^2 e^{\chi_2 }-16 r^2 \chi_3  e^{\chi_2 }+16 r^2
\chi_3 +20 \mathrm{s}^2 \chi_3 e^{\chi_2 }\\\label{g19a} &+48 \pi
r^2+48 \pi \mathrm{s}^2 e^{\chi_2 }\big],\\\nonumber
P_\bot&=\frac{e^{-\chi_2}}{48r^4(\chi_3+2\pi)(\chi_3+4\pi)}\big[r^4
(5 \chi_3 +12 \pi ) \chi_1 '^2+r^3 \chi_1 ' \big(8 (\chi_3 +3 \pi )
\\\nonumber &-r(5 \chi_3 +12 \pi ) \chi_2 '\big)-2 \big\{2 r^3
(\chi_3 +6 \pi ) \chi_2 '-r^4 (5 \chi_3 +12 \pi ) \chi_1''+4 r^2
\chi_3\\\label{g20a} &-4 r^2 \chi_3 e^{\chi_2 } +24 \pi \mathrm{s}^2
e^{\chi_2 }+14 \mathrm{s}^2 \chi_3 e^{\chi_2 }\big\}\big].
\end{align}
Furthermore, the second set (for $\gamma=1$) encodes the effect of
an additional source ($\mathfrak{C}^{\beta}_{\alpha}$) given by
\begin{align}\label{g21}
&8\pi\mathfrak{C}_{0}^{0}=\frac{1}{r}\bigg(\bar{\mathcal{F}}'+\frac{\bar{\mathcal{F}}}{r}\bigg),\\\label{g22}
&8\pi\mathfrak{C}_{1}^{1}=\frac{\bar{\mathcal{F}}}{r}\bigg(\chi_1'+\frac{1}{r}\bigg),\\\label{g23}
&8\pi\mathfrak{C}_{2}^{2}=\frac{\bar{\mathcal{F}}}{4}\bigg(2\chi_1''+\chi_1'^2+\frac{2\chi_1'}{r}\bigg)
+\frac{\bar{\mathcal{F}}'}{2}\bigg(\frac{\chi_1'}{2}+\frac{1}{r}\bigg).
\end{align}

Since we follow the MGD scheme, the energy exchange between both
(seed and additional) matter distributions is not allowed and hence,
they conserve individually. The decoupling of the system
\eqref{g8}-\eqref{g10} into two sectors has successfully been done.
It is observed that there are six unknowns
($\mu,P_r,P_\bot,\mathrm{s},\chi_1,\chi_2$) in the first set of
equations \eqref{g18a}-\eqref{g20a}, therefore, we need a
well-behaved solution and a known form of the electric charge to
continue our analysis. On the other hand, the second sector
\eqref{g21}-\eqref{g23} contains four unknowns
($\bar{\mathcal{F}},\mathfrak{C}_{0}^{0},\mathfrak{C}_{1}^{1},\mathfrak{C}_{2}^{2}$)
and remains the same as that of an uncharged scenario. In this
regard, the only constraint will be needed on $\mathfrak{C}$-sector
to close the system. Following equations help to detect the
effective matter determinants as
\begin{equation}\label{g13}
\bar{\mu}=\mu-\gamma\mathfrak{C}_{0}^{0},\quad
\bar{P}_{r}=P_r+\gamma\mathfrak{C}_{1}^{1}, \quad
\bar{P}_{\bot}=P_\bot+\gamma\mathfrak{C}_{2}^{2},
\end{equation}
and the total anisotropy produced in the system becomes
\begin{equation}\label{g14}
\bar{\Pi}=\bar{P}_{\bot}-\bar{P}_{r}=(P_\bot-P_r)+\gamma(\mathfrak{C}_{2}^{2}-\mathfrak{C}_{1}^{1})=\Pi+\Pi_{\mathfrak{C}},
\end{equation}
where $\Pi$ and $\Pi_{\mathfrak{C}}$ being the anisotropy of the
seed and additional sources, respectively.

\section{Isotropization of Compact Sources}

Since $\bar{\Pi}$ shows the total anisotropy of the spherical
geometry triggered by both seed and extra matter sources which may
be different from that of caused only by the initial source
$\mathbb{T}_{\beta\alpha}$, i.e., $\Pi$. This section discusses the
conversion of the anisotropic system into an isotropic geometry when
a new matter source is included. This can mathematically be
justified by $\bar{\Pi}=0$. We shall show in the following lines
that how the variation in the decoupling parameter affects (or
controls) such evolutionary change. It is observed that the system
is anisotropic for $\gamma=0$ and isotropic when $\gamma=1$. Since
we discuss the later case that yields from Eq.\eqref{g14} as
\begin{equation}\label{g14a}
\Pi_{\mathfrak{C}}=-\Pi \quad \Rightarrow \quad
\mathfrak{C}_{2}^{2}-\mathfrak{C}_{1}^{1}=P_r-P_\bot.
\end{equation}
This condition has been widely used to make the system isotropic
from being anisotropic through minimal/extended decoupling schemes
\cite{37k,39}. A particular acceptable ansatz is now considered in
the following so that the number of unknowns can be decreased. This
is given by
\begin{align}\label{g33}
\chi_1(r)&=\ln\bigg\{\mathcal{C}_2^2\bigg(1+\frac{r^2}{\mathcal{C}_1^2}\bigg)\bigg\},\\\label{g34}
\chi_5(r)&=e^{-\chi_2}=\frac{\mathcal{C}_1^2+r^2}{\mathcal{C}_1^2+3r^2},
\end{align}
leading the matter variables to
\begin{align}\nonumber
\mu&=\frac{1}{12r^4\big(\chi_3+2\pi\big)
\big(\chi_3+4\pi\big)\big(\mathcal{C}_1^2+3r^2\big)^2}\big[3\mathcal{C}_1^2
r^2 \big(\chi_3  \big(9 r^2-10\mathrm{s}^2\big)\\\label{g35} &+24
\pi \big(r^2-\mathrm{s}^2\big)\big)-\mathcal{C}_1^4 \mathrm{s}^2 (5
\chi_3 +12 \pi )+3 r^4 (5 \chi_3 +12 \pi ) \big(2 r^2-3
\mathrm{s}^2\big)\big],\\\nonumber
P_r&=\frac{1}{12r^4\big(\chi_3+2\pi\big)\big(\chi_3+4\pi\big)\big(\mathcal{C}_1^2+3r^2\big)^2}\big[3\mathcal{C}_1^2
r^2 \big(3 r^2 \chi_3 +10 \mathrm{s}^2 \chi_3 \\\label{g36} &+24 \pi
\mathrm{s}^2\big)+\mathcal{C}_1^4 \mathrm{s}^2 (5 \chi_3 +12 \pi )+3
r^4 \big(2 r^2 \chi_3 +15 \mathrm{s}^2 \chi_3+36\pi
\mathrm{s}^2\big)\big],\\\nonumber
P_\bot&=\frac{1}{12r^4\big(\chi_3+2\pi\big)\big(\chi_3+4\pi\big)\big(\mathcal{C}_1^2+3r^2\big)^2}\big[3
\mathcal{C}_1^2 r^2 \big(3 r^2 \chi_3 -14 \mathrm{s}^2 \chi_3
\\\label{g37} &-24 \pi \mathrm{s}^2\big)-\mathcal{C}_1^4 \mathrm{s}^2 (7 \chi_3 +12
\pi )+3 r^4 \big(\chi_3 \big(8 r^2-21 \mathrm{s}^2\big)+12 \pi
\big(r^2-3 \mathrm{s}^2\big)\big)\big],
\end{align}
where $\mathcal{C}_1^2$ and $\mathcal{C}_2^2$ are unknowns that
shall later be determined through continuity of the metric
functions. Some clusters of different particles move in randomly
oriented (circular-like) orbits in their gravitational field. The
above considered solution \eqref{g33} and \eqref{g34} helps to
determine such fields \cite{42a1}. Casadio et al. \cite{37k} also
used them to formulate the decoupled anisotropic solution in
$\mathbb{GR}$.

An important topic of all time for researchers is the junction
conditions which help them to analyze several significant
characteristics of celestial structures at the surface of a sphere,
i.e., $\Sigma:r=\mathcal{R}$. We need to consider an exterior metric
representing vacuum spacetime so that the smooth matching can be
done. Since our considered model \eqref{g61} is equivalent to
$\mathbb{GR}$ in a vacuum, we take the exterior solution provided by
the Reissner-Nordstr\"{o}m line element given by
\begin{equation}\label{g25}
ds^2=-\left(1-\frac{2\bar{\mathcal{M}}}{r}+\frac{\bar{\mathcal{S}}^2}{r^2}\right)dt^2
+\frac{1}{\left(1-\frac{2\bar{\mathcal{M}}}{r}+\frac{\bar{\mathcal{S}}^2}{r^2}\right)}dr^2+
r^2d\vartheta^2+r^2\sin^2\vartheta d\varphi^2,
\end{equation}
where $\bar{\mathcal{S}}$ and $\bar{\mathcal{M}}$ being the total
charge and mass, respectively. The constants $\mathcal{C}_1^2$ and
$\mathcal{C}_2^2$ are obtained as
\begin{eqnarray}\label{g37a}
\mathcal{C}_1^2&=&\frac{\mathcal{R}^2\big(2\mathcal{R}^2-6\bar{\mathcal{M}}\mathcal{R}+3\bar{\mathcal{S}}^2\big)}
{2\bar{\mathcal{M}}\mathcal{R}-\bar{\mathcal{S}}^2},\\\label{g38}
\mathcal{C}_2^2&=&\frac{2\mathcal{R}^2-6\bar{\mathcal{M}}\mathcal{R}+3\bar{\mathcal{S}}^2}{2\mathcal{R}^2}.
\end{eqnarray}

We consider a particular compact star $4U~1820-30$ with a radius
$\mathcal{R}=9.1 \pm 0.4 km$ and mass $\bar{\mathcal{M}}=1.58 \pm
0.06 M_{\bigodot}$ ($M_{\bigodot}$ being mass of the sun) to
graphically analyze the physical characteristics of the anisotropic
models \cite{42aa}. This is a unique compact X-ray binary in the
globular cluster NGC 6624 that exhibits significant luminosity
variability and an extremely short orbital period of 685$s$. Joining
together the constraint \eqref{g14a}, a known form of the charge
\big(involving a constant $\chi_6$ \cite{37abc}\big), the modified
field equations and the considered metric potentials
\eqref{g33}-\eqref{g34} result in a first-order differential
equation as
\begin{align}\nonumber
&r \big(\mathcal{C}_1^2+r^2\big) \big\{8 \pi  r^3
\big(\mathcal{C}_1^2+r^2\big) \big(2 \chi_6 \mathcal{C}_1^4+12
\chi_6  \mathcal{C}_1^2 r^2+18 \chi_6 r^4-3\big)\\\nonumber
&-(\chi_3 +4 \pi ) \big(\mathcal{C}_1^2+2 r^2\big)
\big(\mathcal{C}_1^2+3 r^2\big)^2 \mathcal{F}'(r)\big\}+2 (\chi_3\
+4 \pi )\\\label{g39} &\times \big(\mathcal{C}_1^4+2 \mathcal{C}_1^2
r^2+2 r^4\big) \big(\mathcal{C}_1^2+3 r^2\big)^2\mathcal{F}(r)=0.
\end{align}
Integrating the above equation for the deformation function
$\mathcal{F}(r)$, we get
\begin{equation}\label{g40}
\mathcal{F}(r)=\frac{r^2\big(\mathcal{C}_1^2+r^2\big)}{\mathcal{C}_1^2+2r^2}\bigg[\mathbb{C}_1
+\frac{8\pi}{(\chi_3+4\pi)}\bigg\{\frac{\chi_6\mathcal{C}_1^2}{3}+\frac{1}{2\mathcal{C}_1^2+6r^2}+\chi_6r^2\bigg\}\bigg],
\end{equation}
with $\mathbb{C}_1$ being a real-valued constant. Equation
\eqref{g40} produces the deformed $g_{rr}$ component as
\begin{align}\nonumber
e^{\chi_2}=\chi_5^{-1}&=\big[\big(\mathcal{C}_1^2+r^2\big) \big\{3
\mathcal{C}_1^2 \big(\chi _3+4 \pi \big) \big(\gamma\mathbb{C}_1
r^2+1\big)+3 r^2 \big(\chi _3 \big(3 \gamma\mathbb{C}_1
r^2+2\big)\\\nonumber &+4 \pi \big(3 \gamma\mathbb{C}_1 r^2+\gamma
+2\big)\big)+8 \pi \gamma r^2 \chi _6
\big(\mathcal{C}_1^2+3r^2\big)^2\big\}\big]^{-1}\big[3 \big(\chi
_3+4 \pi \big)\\\label{g40a} &\times \big(5 \mathcal{C}_1^2
r^2+\mathcal{C}_1^4+6 r^4\big)\big].
\end{align}
Hence, the following line element represents minimally deformed
anisotropic solution to the Eqs.\eqref{g8}-\eqref{g10} given by
\begin{equation}\label{g41}
ds^2=-\mathcal{C}_2^2\bigg(1+\frac{r^2}{\mathcal{C}_1^2}\bigg)dt^2+\frac{\mathcal{C}_1^2+3r^2}
{\mathcal{C}_1^2+r^2+\alpha\mathcal{F}\big(\mathcal{C}_1^2+3r^2\big)}dr^2+
r^2d\vartheta^2+r^2\sin^2\vartheta d\varphi^2,
\end{equation}
and the respective state determinants are presented as
\begin{align}\nonumber
\bar{\mu}&=\frac{-1}{24\pi\big(\chi_3+2\pi\big)\big(\chi_3+4\pi\big)\big(\mathcal{C}_1^2+2r^2\big)^2
\big(\mathcal{C}_1^2+3r^2\big)^2}\\\nonumber
&\times\big[\mathcal{C}_1^8 \big\{9 \gamma\mathbb{C}_1 \chi _3^2+2
\pi \chi _3 \big(27 \gamma\mathbb{C}_1+5 (32 \gamma +1) r^2 \chi
_6\big)+8 \pi ^2 \big((80 \gamma +3) r^2 \chi _6\\\nonumber &+9
\gamma\mathbb{C}_1\big)\big\}+\mathcal{C}_1^6 \big\{75
\gamma\mathbb{C}_1 r^2 \chi _3^2+2 \pi \chi _3 \big(9 \big(25 \gamma
\mathbb{C}_1 r^2+2 \gamma -3\big)+2 (408 \gamma +25)
\\\nonumber &\times r^4 \chi _6\big)+24 \pi ^2 \big(25 \gamma\mathbb{C}_1
r^2+3 \gamma +2 (68 \gamma +5) r^4 \chi
_6-6\big)\big\}+\mathcal{C}_1^4 r^2 \big\{2 \pi \chi _3\\\nonumber
&\times \big(675 \gamma\mathbb{C}_1 r^2+60 \gamma +(1992 \gamma
+185) r^4 \chi _6-138\big)+225 \gamma\mathbb{C}_1 r^2 \chi _3^2+24
\pi ^2\\\nonumber &\times  \big(5 \big(15 \gamma\mathbb{C}_1 r^2+2
\gamma -6\big)+(332 \gamma +37) r^4 \chi _6\big)\big\}+3
\mathcal{C}_1^2 r^4 \big\{99 \gamma\mathbb{C}_1 r^2 \chi _3^2+2 \pi
\chi _3\\\nonumber &\times \big(9 \gamma \big(33\mathbb{C}_1
r^2+2\big)+20 (39 \gamma +5) r^4 \chi _6-76\big)+24 \pi ^2
\big(\gamma \big(33\mathbb{C}_1 r^2+3\big)+10
\\\nonumber &\times (13 \gamma +2) r^4 \chi _6-16\big)\big\}+6 r^6
\big\{27 \gamma\mathbb{C}_1 r^2 \chi _3^2+2 \pi \chi _3 \big(81
\gamma\mathbb{C}_1 r^2+6 \gamma +30r^4 \chi _6
\\\nonumber &\times(6 \gamma +1) -20\big)+24 \pi ^2 \big(9 \gamma\mathbb{C}_1 r^2+\gamma
+6 (5 \gamma +1) r^4 \chi _6-4\big)\big\}+24 \pi \gamma
\mathcal{C}_1^{10}\\\label{g46} &\times \big(\chi _3+2 \pi \big)
\chi _6\big],\\\nonumber
\bar{P}_{r}&=\frac{1}{24\pi\big(\chi_3+2\pi\big)\big(\chi_3+4\pi\big)\big(\mathcal{C}_1^2+2r^2\big)
\big(\mathcal{C}_1^2+3r^2\big)^2}\\\nonumber &\times\big[\big(\chi
_3+2 \pi \big) \big(\mathcal{C}_1^2+3 r^2\big)^2 \big\{3 \gamma
\big(\mathbb{C}_1 \mathcal{C}_1^2 \big(\chi _3+4 \pi
\big)+3\mathbb{C}_1 r^2 \chi _3+4 \big(3 \pi\mathbb{C}_1 r^2+\pi
\big)\big)\\\nonumber &+8 \pi \gamma \chi _6 \big(\mathcal{C}_1^2+3
r^2\big)^2\big\}+2 \pi  \big(\mathcal{C}_1^2+2 r^2\big) \big\{r^2
\big(5 \chi _3+12 \pi \big) \chi _6 \big(\mathcal{C}_1^2+3
r^2\big)^2\\\label{g47} &+3 \chi _3 \big(3 \mathcal{C}_1^2+2
r^2\big)\big\}\big],\\\nonumber
\bar{P}_{\bot}&=\frac{1}{24\pi\big(\chi_3+2\pi\big)\big(\chi_3+4\pi\big)\big(\mathcal{C}_1^2+2r^2\big)
\big(\mathcal{C}_1^2+3r^2\big)^2}\\\nonumber
&\times\big[\mathcal{C}_1^6 \big\{3 \gamma\mathbb{C}_1 \big(\chi
_3+2 \pi \big) \big(\chi _3+4 \pi \big)+2 \pi r^2 \chi _6 \big((60
\gamma -7) \chi _3+12 \pi (10 \gamma -1)\big)\big\}\\\nonumber
&+\mathcal{C}_1^4 \big\{27 \gamma\mathbb{C}_1 r^2 \chi _3^2+2 \pi
\chi _3 \big(81 \gamma\mathbb{C}_1 r^2+6 \gamma +8 (39 \gamma -7)
r^4 \chi _6+9\big)+24 \pi ^2\\\nonumber &\times \big(9
\gamma\mathbb{C}_1 r^2+\gamma +4 (13 \gamma -2) r^4 \chi
_6\big)\big\}+3 \mathcal{C}_1^2 r^2 \big\{27 \gamma\mathbb{C}_1 r^2
\chi _3^2+2 \pi \chi _3 \big(81 \gamma\mathbb{C}_1 r^2\\\nonumber
&+6 \gamma +(228 \gamma -49) r^4 \chi _6+14\big)+24 \pi ^2 \big(9
\gamma\mathbb{C}_1 r^2+\gamma +(38 \gamma -7) r^4
\chi_6+1\big)\big\}\\\nonumber &+3 r^4 \big\{27 \gamma\mathbb{C}_1
r^2 \chi _3^2+2 \pi \chi _3 \big(81 \gamma\mathbb{C}_1 r^2+6 \gamma
+6 (30 \gamma -7) r^4 \chi _6+16\big)+24 \pi ^2\\\label{g48} &\times
\big(9 \gamma\mathbb{C}_1 r^2+\gamma +6 (5 \gamma -1) r^4 \chi
_6+2\big)\big\}+8 \pi \gamma \mathcal{C}_1^8 \big(\chi _3+2 \pi
\big) \chi _6\big].
\end{align}

Moreover, the pressure anisotropy corresponding to metric
\eqref{g41} is
\begin{eqnarray}\label{g49}
\bar{\Pi}&=&\frac{r^2\big\{3-2\chi_6\big(\mathcal{C}_1^2+3r^2\big)^2\big\}\big(1-\gamma\big)}{2\big(\chi_3+4\pi\big)
\big(\mathcal{C}_1^2+3r^2\big)^2},
\end{eqnarray}
from which we clearly observe that this factor disappears when
$\gamma=1$, hence, the considered matter source becomes isotropic
for this parametric value. Equations \eqref{g46}-\eqref{g49} provide
the exact solution of $f(\mathbb{R},\mathbb{T})$ field equations
where $\gamma$ belongs to $[0,1]$. In this way, we can thoroughly
follow the process of isotropizing the anisotropic fluid by taking
variation in the decoupling parameter.

\section{Complexity of Compact Sources}

The notion of complexity was initially formulated for a static
sphere \cite{37g} and then extended to the framework of dissipative
dynamical matter configuration \cite{37h}. The basic idea to
establish its definition is that there does not exist any factor
that produces complexity in uniform/isotropic distribution. A
specific factor that measures the complexity of a spacetime
structure (i.e., $\mathbb{Y}_{TF}$) is one of the four different
scalars resulting from the orthogonal decomposition of the curvature
tensor. By following Herrera's technique, we obtain this scalar
factor possessing $f(\mathbb{R},\mathbb{T})$ corrections in the
following form
\begin{equation}\label{g51}
\mathbb{Y}_{TF}(r)=\big(1+\chi_3\big)\bigg(8\pi\Pi+\frac{2\mathrm{s}^2}{r^4}\bigg)
-\frac{4\pi}{r^3}\int_0^r\hat{r}^3\mu'(\hat{r})d\hat{r}.
\end{equation}
The presence of an additional source in the current setup
\eqref{g8}-\eqref{g10} makes the above complexity factor as
\begin{eqnarray}\nonumber
\bar{\mathbb{Y}}_{TF}(r)&=&\big(1+\chi_3\big)\bigg(8\pi\bar{\Pi}+\frac{2\mathrm{s}^2}{r^4}\bigg)
-\frac{4\pi}{r^3}\int_0^r\hat{r}^3\bar{\mu}'(\hat{r})d\hat{r}\\\nonumber
&=&\big(1+\chi_3\big)\bigg(8\pi\Pi+\frac{2\mathrm{s}^2}{r^4}\bigg)
-\frac{4\pi}{r^3}\int_0^r\hat{r}^3\mu'(\hat{r})d\hat{r}\\\label{g54}
&+&8\pi\Pi_{\mathfrak{C}}\big(1+\chi_3\big)+\frac{4\pi}{r^3}\int_0^r\hat{r}^3\mathfrak{C}{_0^0}'(\hat{r})d\hat{r}.
\end{eqnarray}
We can analogously write the above equation as
\begin{eqnarray}\label{g55}
\bar{\mathbb{Y}}_{TF}=\mathbb{Y}_{TF}+\mathbb{Y}_{TF}^{\mathfrak{C}},
\end{eqnarray}
where the sets \eqref{g18a}-\eqref{g20a} and \eqref{g21}-\eqref{g23}
being represented by $\mathbb{Y}_{TF}$ and
$\mathbb{Y}_{TF}^{\mathfrak{C}}$, respectively. Since the model
\eqref{g46}-\eqref{g49} is established for the constraint
$\bar{\Pi}=0$, Eq.\eqref{g54} thus yields
\begin{eqnarray}\label{g56}
\bar{\mathbb{Y}}_{TF}&=&\frac{2\mathrm{s}^2}{r^4}\big(1+\chi_3\big)
-\frac{4\pi}{r^3}\int_0^r\hat{r}^3\bar{\mu}'(\hat{r})d\hat{r}.
\end{eqnarray}
Eq.\eqref{g56} produces the following complexity factor after
combining with the metric \eqref{g41} and the modified field
equations as
\begin{align}\nonumber
\bar{\mathbb{Y}}_{TF}&=2\chi_6r^2\big(1+\chi_3\big)
+\frac{1}{30r^3\big(\chi_3+2\pi\big)\big(\chi_3+4\pi\big)\big(\mathcal{C}_1^2+2r^2\big)^2
\big(\mathcal{C}_1^2+3r^2\big)^2}\\\nonumber &\times\bigg[\pi
\mathcal{C}_1^8 r \big\{5 \chi_3  \big(4 r^4 (8 \chi_6  \gamma
+\chi_6 )+3\big)+16 \pi \chi_6  (20 \gamma +3) r^4\big\}+5
\mathcal{C}_1^6 \big\{48 \pi ^2\\\nonumber &\times \big(2 r^7 (6
\chi_6 \gamma +\chi_6 )-\gamma\mathbb{C}_1 r^5\big)-6 \gamma
\mathbb{C}_1r^5 \chi_3 ^2+\pi r^3 \chi_3 \big(27-36
\gamma\mathbb{C}_1r^2+8 \chi_6\\\nonumber &\times (36 \gamma +5)
r^4\big)\big\}+4 \mathcal{C}_1^4 r^5 \big\{-45 \gamma\mathbb{C}_1r^2
\chi_3^2+5\pi\chi_3\big(24(2-\gamma)-54\gamma\mathbb{C}_1r^2\\\nonumber
&+\chi_6 (240 \gamma +37) r^4\big)+12 \pi ^2 \big(30-20 \gamma -30
\gamma\mathbb{C}_1 r^2+\chi_6 (200 \gamma +37)
r^4\big)\big\}\\\nonumber &-5 \sqrt{3} \pi \sqrt{\mathcal{C}_1^2}
\chi_3 \big(\mathcal{C}_1^4+5\mathcal{C}_1^2r^2+6 r^4\big)^2 \tan
^{-1}\big(\frac{\sqrt{3}r}{\sqrt{\mathcal{C}_1^2}}\big)+30\mathcal{C}_1^2r^7
\big\{2 \pi \chi_3\\\nonumber &\times \big(45-24 \gamma -27
\gamma\mathbb{C}_1 r^2+20 r^4 (6 \chi_6 \gamma +\chi_6 )\big)-9
\gamma \mathbb{C}_1r^2 \chi_3 ^2+24 \pi ^2 \big(8-4 \gamma
\\\nonumber &-3 \gamma\mathbb{C}_1r^2+4 r^4 (5 \chi_6 \gamma +\chi_6
)\big)\big\}+48 \pi r^9 \big\{5 \chi_3 \big(10-3 \gamma +3 r^4 (6
\chi_6 \gamma +\chi_6 )\big)\\\label{g56a} &+6 \pi \big(6 r^4 (5
\chi_6 \gamma +\chi_6 )-5 (\gamma -4)\big)\big\}\bigg].
\end{align}

\subsection{Two Systems with the Same Complexity Factor}

In this subsection, we consider that the scalar factor
$\mathbb{Y}_{TF}$ representing the complexity of an initial matter
source does not possess any change when a new gravitational source
$\mathfrak{C}_{\beta\alpha}$ is added, i.e.,
$\mathbb{Y}_{TF}^{\mathfrak{C}}=0$. This constraint results in
$\bar{\mathbb{Y}}_{TF}=\mathbb{Y}_{TF}$ or
\begin{eqnarray}\label{g56b}
8\pi\Pi_{\mathfrak{C}}\big(1+\chi_3\big)=-\frac{4\pi}{r^3}\int_0^r\hat{r}^3\mathfrak{C}{_0^0}'(\hat{r})d\hat{r}.
\end{eqnarray}
We use Eq.\eqref{g21} to manipulate the right side of \eqref{g56b}
as
\begin{eqnarray}\label{g56c}
-\frac{4\pi}{r^3}\int_0^r\hat{r}^3\mathfrak{C}{_0^0}'(\hat{r})d\hat{r}=
-\frac{1}{2r}\bigg(\mathcal{F}'-\frac{2\mathcal{F}}{r}\bigg),
\end{eqnarray}
yielding the condition \eqref{g56b} as the following first-order
differential equation given by
\begin{eqnarray}\nonumber
&&(\chi_3+1)\bigg\{\mathcal{F}'(r)\bigg(\frac{\chi_1'}{4}+\frac{1}{2r}\bigg)+\mathcal{F}(r)\bigg(\frac{\chi_1''}{2}-\frac{1}{r^2}
+\frac{\chi_1'^2}{4}-\frac{\chi_1'}{2r}\bigg)\bigg\}\\\label{g56d}
&&+\frac{1}{2r}\bigg(\mathcal{F}'-\frac{2\mathcal{F}}{r}\bigg)=0.
\end{eqnarray}
This equation possesses the temporal metric function describing the
original anisotropic matter field, hence increases the unknowns. The
Tolman IV ansatz is used in this regard to make the system solvable
as
\begin{align}\label{g57}
\chi_1(r)&=\ln\bigg\{\mathcal{C}_2^2\bigg(1+\frac{r^2}{\mathcal{C}_1^2}\bigg)\bigg\},\\\label{g58}
\chi_5(r)&=e^{-\chi_2}=\frac{\big(\mathcal{C}_1^2+r^2\big)
\big(\mathcal{C}_3^2-r^2\big)}{\mathcal{C}_3^2\big(\mathcal{C}_1^2+2r^2\big)},
\end{align}
triggered by the density and pressure isotropy as
\begin{align}\nonumber
\mu&=\frac{1}{12\mathcal{C}_3^2\big(\chi_3+2\pi\big)\big(\chi_3+4\pi\big)\big(\mathcal{C}_1^2+2r^2\big)^2}\\\nonumber
&\times\big[\mathcal{C}_1^4 \big\{12 (\chi_3 +3 \pi )-\chi_6
\mathcal{C}_3^2 r^2 (5 \chi_3 +12 \pi )\big\}+\mathcal{C}_1^2
\big\{12 r^2 (2 \chi_3 +7 \pi )-\mathcal{C}_3^2\\\nonumber &\times
(5 \chi_3 +12 \pi ) \big(4 \chi_6 r^4-3\big)\big\}+2 r^2 \big\{9 r^2
(\chi_3 +4 \pi )-2 \mathcal{C}_3^2 \big(5 \chi_6  r^4 \chi_3 -3
\chi_3\\\label{g59} &+12 \pi \chi_6 r^4 -6 \pi
\big)\big\}\big],\\\nonumber
P&=\frac{1}{12\mathcal{C}_3^2\big(\chi_3+2\pi\big)\big(\chi_3+4\pi\big)\big(\mathcal{C}_1^2+2r^2\big)^2}\\\nonumber
&\times\big[\mathcal{C}_1^4 \big\{\chi_6  \mathcal{C}_3^2 r^2 (5
\chi_3 +12 \pi )-12 \pi \big\}+\mathcal{C}_1^2 \big\{\mathcal{C}_3^2
\big(20 \chi_6  r^4 \chi_3 +48 \pi \chi_6 r^4 +12 \pi\\\nonumber &+9
\chi_3 \big)-12 r^2 (\chi_3 +5 \pi )\big\}+2 r^2 \big(2
\mathcal{C}_3^2 \big\{5 \chi_6 r^4 \chi_3 +12 \pi  \chi_6 r^4+3
\chi_3 +6 \pi \big\}\\\label{g60} &-9 r^2 (\chi_3 +4 \pi)\big)\big].
\end{align}
Equations \eqref{g37a} and \eqref{g38} supply the unknowns
$\mathcal{C}_1^2$ and $\mathcal{C}_2^2$, whereas $\mathcal{C}_3^2$
has the following form
\begin{align}\label{g60a}
\mathcal{C}_3^2&=\frac{2\mathcal{R}^3\big(\bar{\mathcal{S}}^2-2\bar{\mathcal{M}}\mathcal{R}+\mathcal{R}^2\big)}
{2\bar{\mathcal{M}}\bar{\mathcal{S}}-4\bar{\mathcal{M}}^2\mathcal{R}+\bar{\mathcal{S}}\mathcal{R}+2\bar{\mathcal{M}}\mathcal{R}^2}.
\end{align}
Insertion of $g_{tt}$ metric component \eqref{g57} in
Eq.\eqref{g56d} gives
\begin{equation}\label{g60b}
\mathcal{F}(r)=\frac{\mathbb{C}_2r^2\big(\mathcal{C}_1^2+r^2\big)}{\mathcal{C}_1^2\big(2+\chi_3\big)+r^2\big(2\chi_3+3\big)},
\end{equation}
involving $\mathbb{C}_2$ as a constant of integration. Consequently,
the deformed expression of $g_{rr}$ component becomes
\begin{align}\label{g60c}
e^{\chi_2}&=\chi_5^{-1}=\frac{\mathcal{C}_3^2\big(\mathcal{C}_1^2+2r^2\big)}{\big(\mathcal{C}_1^2+r^2\big)}\bigg\{\mathcal{C}_3^2
+r^2\bigg(\frac{\gamma\mathbb{C}_2\mathcal{C}_3^2\big(\mathcal{C}_1^2+2r^2\big)}{\mathcal{C}_1^2\big(2+\chi_3\big)
+r^2\big(2\chi_3+3\big)}-1\bigg)\bigg\}^{-1},
\end{align}
and the total complexity factor $\bar{\mathbb{Y}}_{TF}$ defined in
Eq.\eqref{g54} leads to
\begin{align}\nonumber
\bar{\mathbb{Y}}_{TF}&=\mathbb{Y}_{TF}=\frac{1}{16r^3\mathcal{C}_3^2\big(\chi_3+2\pi\big)
\big(\chi_3+4\pi\big)\big(\mathcal{C}_1^2+2r^2\big)^2}\bigg[2 r
\big\{45 \pi\mathcal{C}_1^6 \chi_3 \\\nonumber &+2\mathcal{C}_1^4
\big(120 \chi_6 r^4\mathcal{C}_3^2 \chi_3 ^2 (\chi_3 +1)+96 \pi ^2
\chi_6 r^4 \mathcal{C}_3^2+5 \pi \chi_3  \big(8  (6 \chi_3
+7)\\\nonumber &\times \chi_6 r^4\mathcal{C}_3^2+15
r^2+9\mathcal{C}_3^2\big)\big)+4\mathcal{C}_1^2\big(240 \chi_6
r^6\mathcal{C}_3^2 \chi_3 ^2 (\chi_3 +1)+5 \pi r^2 \chi_3
\\\nonumber &\times \big(16
\chi_6  r^4\mathcal{C}_3^2 (6 \chi_3 +7)+24 r^2+15
\mathcal{C}_3^2\big)+48 \pi ^2 r^4 \big(4 \chi_6  r^2
\mathcal{C}_3^2+5\big)\big)\\\nonumber &+64 r^4 \mathcal{C}_3^2
\big(15 \chi_6 r^4 \chi_3 ^2 (\chi_3 +1)+5 \pi  \chi_3 \big(\chi_6
r^4 (6 \chi_3 +7)+3\big)+6 \pi ^2\\\label{g60d} &\times\big(2 \chi_6
r^4+5\big)\big)\big\}-45 \sqrt{2} \pi  \sqrt{\mathcal{C}_1^2} \chi_3
\big(\mathcal{C}_1^2+2 r^2\big)^2 (\mathcal{C}_1^2+2\mathcal{C}_3^2)
\tan^{-1}\bigg(\frac{\sqrt{2}r}{\sqrt{\mathcal{C}_1^2}}\bigg)\bigg].
\end{align}

\subsection{Constructing Solutions with Null Complexity}

This subsection establishes a solution to the
$f(\mathbb{R},\mathbb{T})$ field equations for a particular
constraint depending on the complexity factor. It is considered that
the original anisotropic source does possess the complexity such
that $\mathbb{Y}_{TF}\neq0$, however, the system becomes free from
complexity once a new gravitational source is added. This can
mathematically be expressed by $\bar{\mathbb{Y}}_{TF}=0$, that leads
Eq.\eqref{g55} in terms of ansatz \eqref{g57} and \eqref{g58} to
\begin{align}\nonumber
&\frac{120r\big(\chi_6+1\big)}{\big(\mathcal{C}_1^2+r^2\big)^2}\big[r\big(\mathcal{C}_1^4+3\mathcal{C}_1^2r^2+2r^4\big)
\mathcal{F}'(r)-2\big(\mathcal{C}_1^4+2\mathcal{C}_1^2r^2+2r^4\big)\mathcal{F}(r)\big]\\\nonumber
&+\frac{1}{\big(\chi_3+2\pi\big)\big(\chi_3+4\pi\big)\mathcal{C}_3^2\big(\mathcal{C}_1^2+2r^2\big)^2}
\bigg[2 r \big(45 \pi  \mathcal{C}_1^6 \chi_3+2 \mathcal{C}_1^4
\big(120 \chi_6  r^4 \mathcal{C}_3^2 \chi_3 ^2 \\\nonumber &\times
(\chi_3 +1)+96 \pi ^2 \chi_6 r^4 \mathcal{C}_3^2+5 \pi \chi_3 \big(8
\chi_6 r^4 \mathcal{C}_3^2 (6 \chi_3 +7)+15 r^2+9
\mathcal{C}_3^2\big)\big)\\\nonumber &+4 \mathcal{C}_1^2 \big(240
\chi_6 r^6 \mathcal{C}_3^2 \chi_3 ^2 (\chi_3 +1)+5 \pi r^2 \chi_3
\big(24 r^2+15\mathcal{C}_3^2+16  (6 \chi_3 +7)\\\nonumber
&\times\chi_6 r^4 \mathcal{C}_3^2\big)+48 \pi ^2 r^4 \big(4 \chi_6
r^2\mathcal{C}_3^2+5\big)\big)+64 r^4\mathcal{C}_3^2\big(15 \chi_6
r^4 \chi_3 ^2 (\chi_3 +1)+5 \pi  \chi_3\\\nonumber
&\times\big(\chi_6 r^4 (6 \chi_3 +7)+3\big)+6 \pi ^2 \big(2 \chi_6
r^4+5\big)\big)\big)-45 \sqrt{2} \pi  \sqrt{\mathcal{C}_1^2} \chi_3
\big(\mathcal{C}_1^2+2 r^2\big)^2\\\label{60e}
&\times(\mathcal{C}_1^2+2\mathcal{C}_3^2)\tan^{-1}\bigg(\frac{\sqrt{2}
r}{\sqrt{\mathcal{C}_1^2}}\bigg)\bigg]-120r\big(2\mathcal{F}(r)-r\mathcal{F}'(r)\big)=0.
\end{align}
The solution belonging to the above equation is
\begin{align}\nonumber
\mathcal{F}(r)&=\frac{r^2\big(\mathcal{C}_1^2+r^2\big)}{120\mathcal{C}_3^2
\big(\chi_3^2+6\pi\chi_3+8\pi^2\big)\big(\chi_3\mathcal{C}_1^2+2\mathcal{C}_1^2+2\chi_3r^2+3r^2\big)}
\bigg[\big(\mathcal{C}_1^2+2\mathcal{C}_3^2\big)\\\nonumber
&\times\bigg\{\frac{30\pi\chi_3}{r^2}+\frac{60\pi(3\chi_3+8\pi)}{\mathcal{C}_1^2+2r^2}
-\frac{15\chi_3\sqrt{2\mathcal{C}_1^2}\tan^{-1}\bigg(\frac{\sqrt{2}
r}{\sqrt{C}_1^2}\bigg)}{r^3}\bigg\}-16\chi_6r^2\\\label{60f}
&\times\big\{12\pi^2+15\chi_3^2(1+\chi_3)+5\pi\chi_3(7+6\chi_3)\big\}\bigg]
+\frac{\mathbb{C}_3r^2\big(\mathcal{C}_1^2+r^2\big)}{\chi_3\mathcal{C}_1^2+2\mathcal{C}_1^2+2\chi_3
r^2+3r^2},
\end{align}
where $\mathbb{C}_3$ being an integration constant with dimension
$\frac{1}{\ell^2}$. The transformation \eqref{g17} modifies the
radial coefficient in terms of the function \eqref{60f} as
\begin{align}\label{g60fa}
e^{\chi_2}&=\chi_5^{-1}=\frac{\mathcal{C}_3^2\big(\mathcal{C}_1^2+2r^2\big)}{\big(\mathcal{C}_1^2+r^2\big)
\big(\mathcal{C}_3^2-r^2\big)+\gamma\mathcal{C}_3^2\mathcal{F}(r)\big(\mathcal{C}_1^2+2r^2\big)}.
\end{align}
Hence, the corresponding matter determinants take the final form as
\begin{align}\nonumber
\bar{\mu}&=\frac{9 \mathcal{C}_1^2 \big(3 \chi _3+8 \pi \big)-r^2
\big(5 \chi _3+12 \pi \big) \chi _6 \big(\mathcal{C}_1^2+3
r^2\big)^2+6 r^2 \big(5 \chi _3+12 \pi \big)}{12 \big(\chi _3+2 \pi
\big) \big(\chi _3+4 \pi \big) \big(\mathcal{C}_1^2+3
r^2\big)^2}\\\nonumber &-\frac{\gamma}{960 \pi \mathcal{C}_3^2
\big(\mathcal{C}_1^2 \big(\chi _3+2\big)+r^2 \big(2 \chi
_3+3\big)\big)^2}\bigg[120 \mathcal{C}_3^2\mathbb{C}_3
\big(\mathcal{C}_1^2 r^2 \big(7 \chi _3+13\big)+3
\\\nonumber &\times \mathcal{C}_1^4\big(\chi _3+2\big)+3 r^4 \big(2 \chi
_3+3\big)\big)+\frac{30 \pi\mathcal{C}_1^2}{r \big(\chi _3+2 \pi
\big) \big(\chi _3+4 \pi \big) \big(\mathcal{C}_1^2+2
r^2\big)^2}\\\nonumber &\times\bigg\{4 \mathcal{C}_1^2 r^5 \big(\chi
_3 \big(14 \chi _3+32 \pi +27\big)+60 \pi \big)+8 \mathcal{C}_1^4
r^3 \big(\chi _3 \big(7 \chi _3+18 \pi +14\big)\\\nonumber &+34 \pi
\big)+4 \sqrt{2} \mathcal{C}_1^3 r^4 \chi _3 \big(\chi _3+1\big)
\tan ^{-1}\bigg(\frac{\sqrt{2} r}{\sqrt{C_1^2}}\bigg)+4 \sqrt{2}
\mathcal{C}_1^5 r^2 \chi _3 \big(\chi _3+1\big) \\\nonumber
&\times\tan ^{-1}\bigg(\frac{\sqrt{2}
r}{\sqrt{\mathcal{C}_1^2}}\bigg)+2 \mathcal{C}_1^6 r \big(\chi _3
\big(9 \chi _3+24 \pi +19\big)+48 \pi \big)+\sqrt{2} \big(\chi
_3+1\big)\\\nonumber &\times \mathcal{C}_1^7 \chi _3 \tan
^{-1}\bigg(\frac{\sqrt{2} r}{\sqrt{\mathcal{C}_1^2}}\bigg)+16 r^7
\big(\chi _3+2 \pi \big) \big(2 \chi _3+3\big)\bigg\}+4
\mathcal{C}_3^2\big\{r \big(\chi _3+2 \pi \big)\\\nonumber
&\times\big(\chi _3+4 \pi \big) \big(\mathcal{C}_1^2+2
r^2\big)^2\big\}^{-1}\bigg\{6 \mathcal{C}_1^6 r \big(5 \pi \big(\chi
_3 \big(9 \chi _3+24 \pi +19\big)+48 \pi \big)\\\nonumber &-2 r^4
\big(11 \chi _3+21\big) \big(5 \chi _3 \big(3 \chi _3 \big(\chi _3+2
\pi +1\big)+7 \pi \big)+12 \pi ^2\big) \chi _6\big)+60 \sqrt{2} \pi
\mathcal{C}_1^3\\\nonumber &\times r^4 \chi _3 \big(\chi _3+1\big)
\tan ^{-1}\bigg(\frac{\sqrt{2} r}{\sqrt{\mathcal{C}_1^2}}\bigg)-20
\mathcal{C}_1^8 r^3 \big(\chi _3+2\big) \big(5 \chi _3 \big(3 \chi
_3 \big(\chi _3+2 \pi +1\big)\\\nonumber &+7 \pi \big)+12 \pi
^2\big) \chi _6+60 \sqrt{2} \pi \mathcal{C}_1^5 r^2 \chi _3
\big(\chi _3+1\big) \tan ^{-1}\bigg(\frac{\sqrt{2}
r}{\sqrt{\mathcal{C}_1^2}}\bigg)+4 \mathcal{C}_1^2 r^5 \big(15
\pi\\\nonumber &\times \big(\chi _3 \big(14 \chi _3+32 \pi
+27\big)+60 \pi \big)-4 r^4 \big(23 \chi _3+38\big) \big(5 \chi _3
\big(3 \chi _3 \big(\chi _3+2 \pi +1\big)\\\nonumber &+7 \pi
\big)+12 \pi ^2\big) \chi _6\big)+4 \mathcal{C}_1^4 r^3 \big(30 \pi
\big(\chi _3 \big(7 \chi _3+18 \pi +14\big)+34 \pi \big)-r^4 \big(82
\chi _3\\\nonumber &+147\big) \big(5 \chi _3 \big(3 \chi _3
\big(\chi _3+2 \pi +1\big)+7 \pi \big)+12 \pi ^2\big) \chi
_6\big)+15 \sqrt{2} \pi  \mathcal{C}_1^7 \chi _3 \big(\chi
_3+1\big)\\\nonumber &\times \tan ^{-1}\bigg(\frac{\sqrt{2}
r}{\sqrt{\mathcal{C}_1^2}}\bigg)+80 r^7 \big(2 \chi _3+3\big) \big(3
\pi \big(\chi _3+2 \pi \big)-r^4 \big( \big(3 \chi _3 \big(\chi _3+2
\pi +1\big)\\\label{60g} &+7 \pi \big)5 \chi _3+12 \pi ^2\big) \chi
_6\big) \bigg\}\bigg],\\\nonumber
\bar{P}_{r}&=\frac{1}{960}\bigg[\frac{80 \big\{r^2 \big(5 \chi _3+12
\pi \big) \chi _6 \big(\mathcal{C}_1^2+3 r^2\big)^2+3 \chi _3 \big(3
\mathcal{C}_1^2+2 r^2\big)\big\}}{\big(\chi _3+2 \pi \big) \big(\chi
_3+4 \pi \big) \big(\mathcal{C}_1^2+3 r^2\big)^2}+\gamma
r^2\big\{\pi\\\nonumber &\times\big(\mathcal{C}_1^2 \big(\chi
_3+2\big)+r^2 \big(2 \chi _3+3\big)\big)\mathcal{C}_3^2 \big(\chi
_3+2\pi\big)\big(\chi_3+4\pi\big)\big\}^{-1}\big(\mathcal{C}_1^2+r^2\big)\\\nonumber
&\times\bigg(\frac{2}{\mathcal{C}_1^2+r^2}+\frac{1}{r^2}\bigg)\bigg\{120
\mathbb{C}_3 \mathcal{C}_3^2 \big(\chi _3+2 \pi \big) \big(\chi _3+4
\pi \big)-15 \sqrt{2} \pi r^{-3}  \mathcal{C}_1  \chi _3\\\nonumber
&\times \big(\mathcal{C}_1^2+2 \mathcal{C}_3^2\big)\tan
^{-1}\bigg(\frac{\sqrt{2} r}{\sqrt{\mathcal{C}_1^2}}\bigg)-16
\mathcal{C}_3^2 r^2 \big(5 \chi _3 \big(3 \chi _3 \big(\chi _3+2 \pi
+1\big)+7 \pi \big)\\\label{60h} &+12 \pi ^2\big) \chi _6+30 \pi
r^{-2} \big(\mathcal{C}_1^2+2 \mathcal{C}_3^2\big) \chi _3+\frac{60
\pi \big(\mathcal{C}_1^2+2 \mathcal{C}_3^2\big) \big(3 \chi _3+8 \pi
\big)}{\mathcal{C}_1^2+2 r^2}\bigg\}\bigg],\\\nonumber
\bar{P}_{\bot}&=\frac{3 \big(\chi _3 \big(3 \mathcal{C}_1^2+8
r^2\big)+12 \pi r^2\big)-r^2 \big(7 \chi _3+12 \pi \big) \chi _6
\big(\mathcal{C}_1^2+3 r^2\big)^2}{12 \big(\chi _3+2 \pi \big)
\big(\chi _3+4 \pi \big) \big(\mathcal{C}_1^2+3 r^2\big)^2}+\gamma
r^2\\\nonumber &\times\big(2\mathcal{C}_1^2+r^2\big)\big\{960 \pi
\mathcal{C}_3^2 \big(\chi _3+2 \pi \big) \big(\chi _3+4 \pi \big)
\big(\mathcal{C}_1^2+r^2\big) \big(\mathcal{C}_1^2 \big(\chi
_3+2\big)\\\nonumber &+r^2 \big(2 \chi
_3+3\big)\big)\big\}^{-1}\bigg\{120 \mathbb{C}_3 \mathcal{C}_3^2
\big(\chi _3+2 \pi \big) \big(\chi _3+4 \pi \big)-15 \sqrt{2} \pi
r^{-3}  \mathcal{C}_1  \chi _3\\\nonumber &\times
\big(\mathcal{C}_1^2+2 \mathcal{C}_3^2\big)\tan
^{-1}\bigg(\frac{\sqrt{2} r}{\sqrt{\mathcal{C}_1^2}}\bigg)-16
\mathcal{C}_3^2 r^2 \big(5 \chi _3 \big(3 \chi _3 \big(\chi _3+2 \pi
+1\big)+7 \pi \big)\\\nonumber &+12 \pi ^2\big) \chi _6+30 \pi
r^{-2} \big(\mathcal{C}_1^2+2 \mathcal{C}_3^2\big) \chi _3+\frac{60
\pi \big(\mathcal{C}_1^2+2 \mathcal{C}_3^2\big) \big(3 \chi _3+8 \pi
\big)}{\mathcal{C}_1^2+2 r^2}\bigg\}\\\nonumber &+\frac{\gamma
r^{-2}\mathcal{C}_3^{-2}}{960 \pi  \big(\mathcal{C}_1^2 \big(\chi
_3+2\big)+r^2 \big(2 \chi _3+3\big)\big)^2 \big(\chi _3+2 \pi \big)
\big(\chi _3+4 \pi \big)\big(\mathcal{C}_1^2+2r^2\big)^2}\\\nonumber
&\times\bigg(\frac{r}{2(\mathcal{C}_1^2+r^2)}+\frac{1}{2r}\bigg)\bigg[240\mathbb{C}_3\mathcal{C}_3^2
r^3\big(2r^2 \mathcal{C}_1^2\big(\chi _3+2\big)+\mathcal{C}_1^4
\big(\chi _3+2\big)+r^4\\\nonumber &\times \big(2 \chi
_3+3\big)\big) \big(\chi _3+2 \pi \big) \big(\chi _3+4 \pi \big)
\big(\mathcal{C}_1^2+2 r^2\big)^2-256 \mathcal{C}_3^2 r^{13} \big(2
\chi _3+3\big)\\\nonumber &\times \big(5 \chi _3 \big(3 \chi _3
\big(\chi _3+2 \pi +1\big)+7 \pi \big)+12 \pi ^2\big) \chi _6+120
\sqrt{2} \pi \mathcal{C}_1 \mathcal{C}_3^2 r^8 \chi _3 \big(2 \chi
_3\\\nonumber &+3\big) \tan ^{-1}\bigg(\frac{\sqrt{2}
r}{\sqrt{\mathcal{C}_1^2}}\bigg)+15 \sqrt{2} \pi \mathcal{C}_1^7 r^2
\chi _3 \big(6 \mathcal{C}_3^2 \big(3 \chi _3+5\big)+13 r^2 \big(2
\chi _3+3\big)\big)\\\nonumber &\times \tan
^{-1}\bigg(\frac{\sqrt{2} r}{\sqrt{\mathcal{C}_1^2}}\bigg)+15
\sqrt{2} \pi \mathcal{C}_1^9 \chi _3 \big(2 \mathcal{C}_3^2
\big(\chi _3+2\big)+3 r^2 \big(3 \chi _3+5\big)\big)\\\nonumber &
\tan ^{-1}\bigg(\frac{\sqrt{2} r}{\sqrt{\mathcal{C}_1^2}}\bigg)+24
\mathcal{C}_1^2 \mathcal{C}_3^2 r^7 \big(-16 r^4 \big(3 \chi
_3+5\big) \big(5 \chi _3\big(3 \chi _3 \big(\chi _3+2 \pi
+1\big)\\\nonumber &+7 \pi \big)+12 \pi ^2\big) \chi _6-5 \pi \chi
_3 \big(6 \chi _3+1\big)+80 \pi ^2\big)+60 \sqrt{2} \pi
\mathcal{C}_1^3 r^6 \chi _3 \big(2 \mathcal{C}_3^2 \big(7 \chi
_3\\\nonumber &+10\big)+r^2 \big(2 \chi _3+3\big)\big) \tan
^{-1}\bigg(\frac{\sqrt{2} r}{\sqrt{\mathcal{C}_1^2}}\bigg)+2
\mathcal{C}_1^8 r \big(15 \pi \big(\chi _3 \big(\big(5 \chi _3+32
\pi +13\big)\\\nonumber &\times r^2-2 \mathcal{C}_3^2 \big(\chi
_3+2\big)\big)+64 \pi r^2\big)-32 \mathcal{C}_3^2 r^4 \big(\chi
_3+2\big) \big(5 \chi _3 \big(3 \chi _3 \big(\chi _3+2 \pi
+1\big)\\\nonumber &+7 \pi \big)+12 \pi ^2\big) \chi _6\big)+30
\sqrt{2} \pi \mathcal{C}_1^5 r^4 \chi _3 \big(13 \mathcal{C}_3^2
\big(2 \chi _3+3\big)+2 r^2 \big(7 \chi _3+10\big)\big)\\\nonumber
&\times \tan ^{-1}\bigg(\frac{\sqrt{2}
r}{\sqrt{\mathcal{C}_1^2}}\bigg)+4 \mathcal{C}_1^4 r^5
\big(\mathcal{C}_3^2 \big(15 \pi \big(\chi _3 \big(8 \chi _3+64 \pi
+27\big)+128 \pi \big)-16 r^4\\\nonumber &\times \big(16 \chi
_3+29\big) \big(5 \chi _3 \big(3 \chi _3 \big(\chi _3+2 \pi
+1\big)+7 \pi \big)+12 \pi ^2\big) \chi _6\big)+15 \pi r^2 \big(16
\pi\\\nonumber & -\chi _3 \big(6 \chi _3+1\big)\big)\big)+2
\mathcal{C}_1^6 r^3 \big(2 \mathcal{C}_3^2 \big(15 \pi \big(\chi _3
\big(5 \chi _3+32 \pi +13\big)+64 \pi \big)-8 r^4\\\nonumber &\times
\big(13 \chi _3+25\big) \big(5 \chi _3 \big(3 \chi _3 \big(\chi _3+2
\pi +1\big)+7 \pi \big)+12 \pi ^2\big) \chi _6\big)+15 \pi  r^2
\big(\chi _3 \\\nonumber &\times \big(8 \chi _3+64 \pi +27\big)+128
\pi \big)\big)-30 \pi \mathcal{C}_1^{10} r \chi _3 \big(\chi
_3+2\big)+15 \sqrt{2} \pi  \mathcal{C}_1^{11} \chi _3 \\\label{60i}
&\times\big(\chi _3+2\big) \tan ^{-1}\bigg(\frac{\sqrt{2}
r}{\sqrt{\mathcal{C}_1^2}}\bigg)\bigg],
\end{align}
and the corresponding anisotropy can be obtained by taking the
difference of Eqs.\eqref{60h} and \eqref{60i}.

\section{Graphical Analysis of the Obtained Models}

A numerical solution of the following first-order differential
equation yields the mass of a spherical fluid configuration by
\begin{equation}\label{g63}
\frac{d\bar{m}(r)}{dr}=4\pi r^2 \bar{\mu},
\end{equation}
where $\bar{\mu}$ (involving $f(\mathbb{R},\mathbb{T})$ corrections)
is presented in Eqs.\eqref{g46} and \eqref{60g} for the first and
second models, respectively. We solve the above equation for an
initial condition $\bar{m}(0)=0$. The particles in any
self-gravitating system arranged in some particular manner which
helps to know the compactness $\big(\tau(r)\big)$ in that body. The
tightness of the particles by which they are ordered in a system can
also be measured with the help of this factor. Moreover, the
mass-radius ratio also helps to estimate $\tau(r)$. A feasible
spherical structure must possess its maximum value by $\frac{4}{9}$
\cite{42a}. The interesting fact about a compactness factor is that
the wavelength of an electromagnetic radiations emit by a celestial
body necessarily affect by it. A sufficient attractive gravitational
force in a compact star changes the path in which those radiations
move. Based on the above discussion, we specify the redshift in the
following manner
\begin{equation}
z(r)=\frac{1-\sqrt{1-2\tau(r)}}{\sqrt{1-2\tau(r)}},
\end{equation}
whose maximum value at the spherical surface is $2$ \big(or
$5.211$\big) for perfect \cite{42a} \big(or anisotropic
\cite{42b}\big) fluids, respectively.

Certain constraints are of great importance in the field of
astronomy to check the physical viability of the obtained models in
any theory of gravity. These bounds also help to check whether there
exists a usual matter in the interior of a compact star or not. They
are termed as the energy conditions in the literature whose
dissatisfaction guarantees the presence of an exotic fluid. The
energy conditions are the linear combinations of different matter
determinants (i.e., density and pressure) that must be fulfilled for
an acceptable model. They take the form for the considered scenario
as
\begin{eqnarray}\nonumber
&&\bar{\mu}+\frac{\mathrm{s}^2}{8\pi r^4} \geq 0, \quad
\bar{\mu}+\bar{P}_{r} \geq 0,\\\nonumber
&&\bar{\mu}+\bar{P}_{\bot}+\frac{\mathrm{s}^2}{4\pi r^4} \geq 0,
\quad \bar{\mu}-\bar{P}_{r}+\frac{\mathrm{s}^2}{4\pi r^4} \geq
0,\\\label{g50} &&\bar{\mu}-\bar{P}_{\bot} \geq 0, \quad
\bar{\mu}+\bar{P}_{r}+2\bar{P}_{\bot}+\frac{\mathrm{s}^2}{4\pi r^4}
\geq 0.
\end{eqnarray}

The stability plays a crucial role in the analysis of the evolution
of self-gravitating models. We shall discuss this phenomenon with
the help of two strategies in the following. Firstly, we define the
sound speeds in radial
$\big(V_{sr}^{2}=\frac{d\bar{P}_{r}}{d\bar{\mu}}\big)$ and
tangential
$\big(V_{s\bot}^{2}=\frac{d\bar{P}_{\bot}}{d\bar{\mu}}\big)$
directions. It was pointed out that the speed of sound in a medium
must be less than that of light to maintain the causality, i.e., $0
< V_{sr}^{2},~ V_{s\bot}^{2} < 1$ \cite{42bb}. Another criterion to
do so has been suggested by Herrera \cite{42ba,42bc}, according to
which the inequality $0 < |V_{s\bot}^{2}-V_{sr}^{2}| < 1$ should
hold to get a physically stable model.
\begin{figure}\center
\epsfig{file=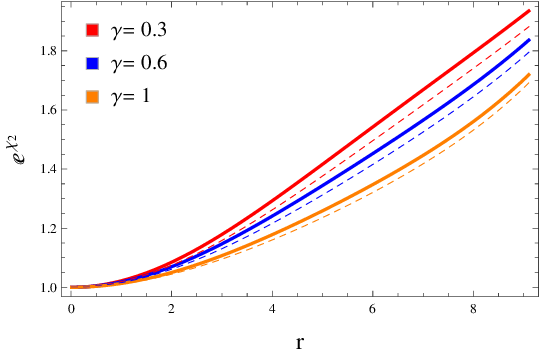,width=0.4\linewidth}
\caption{Deformed $g_{rr}$ metric function \eqref{g40a} versus $r$
(in km) for $\bar{\mathcal{S}}=0.1$ (solid lines) and
$\bar{\mathcal{S}}=0.8$ (dashed lines) corresponding to
$\bar{\Pi}=0$.}
\end{figure}

We adopt the $f(\mathbb{R},\mathbb{T})$ model \eqref{g61} to discuss
the graphical interpretation of the both resulting charged
solutions, their corresponding deformation functions as well as the
complexity factors. For the first solution, different values of the
decoupling parameter and charge along with an integration constant
$\mathbb{C}_{1}=-0.003$ and $\chi_3=0.3$ are chosen so that the
corresponding physical characteristics can be explored. Figure
\textbf{1} assures an acceptable (singularity free and increasing
outwards) behavior of the deformed $g_{rr}$ potential \eqref{g40a}
for $0<r<\mathcal{R}$. A resulting model would be found acceptable
only if the governing state parameters representing interior
distribution (like the energy density and pressure) are maximum
(minimum) and positively finite at the center (boundary) of an
astrophysical body. Figure \textbf{2} depicts such behavior for the
first solution being represented by Eqs.\eqref{g46}-\eqref{g49}. The
upper left plot of the energy density shows its maximum value in the
star's core. However, the increment in the decoupling parameter as
well as an electric charge decreases the density of the fluid. On
the other hand, the pressure in radial and transverse directions
shows the opposite behavior to that of the density corresponding to
the parameter $\gamma$. Additionally, the pressure components
decrease with the increment in an electric charge. The star's
boundary contains only the tangential pressure while the radial one
disappears at that point. Figure \textbf{2} (last plot) also
exhibits the plot of the anisotropic factor that vanishes only at
the core for $\gamma=0.3,~0.6$ and throughout for $\gamma=1$. Hence,
the isotropization of the developed solution is clearly observed
from the graphical analysis.
\begin{figure}\center
\epsfig{file=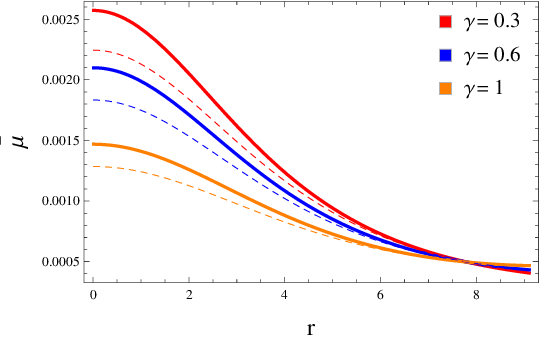,width=0.4\linewidth}\epsfig{file=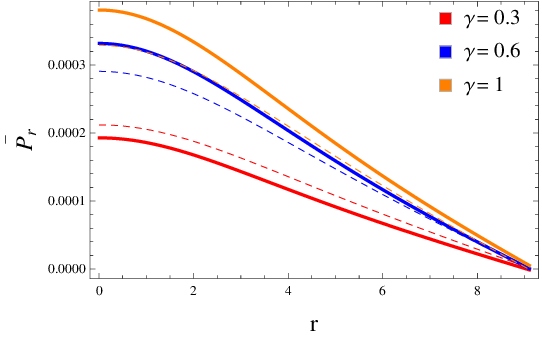,width=0.4\linewidth}
\epsfig{file=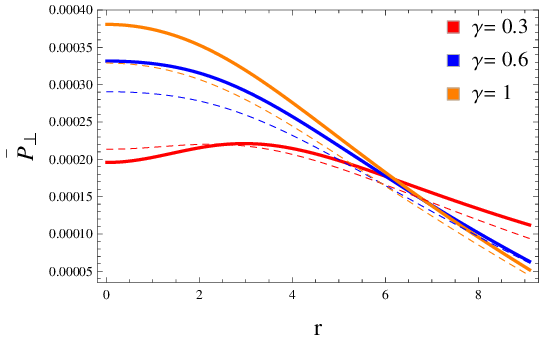,width=0.4\linewidth}\epsfig{file=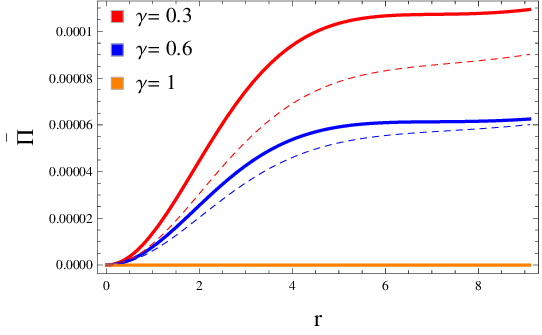,width=0.4\linewidth}
\caption{Matter determinants and anisotropy (in km$^{-2}$) versus
$r$ (in km) for $\bar{\mathcal{S}}=0.1$ (solid lines) and
$\bar{\mathcal{S}}=0.8$ (dashed lines) corresponding to
$\bar{\Pi}=0$.}
\end{figure}
\begin{figure}\center
\epsfig{file=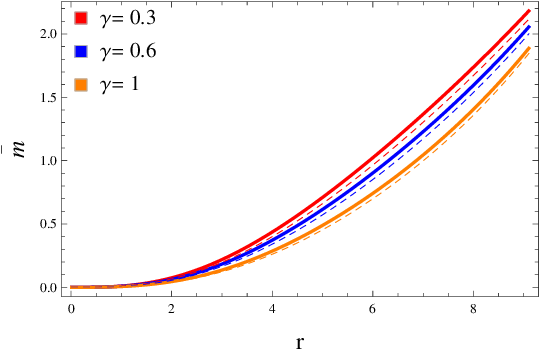,width=0.4\linewidth}\epsfig{file=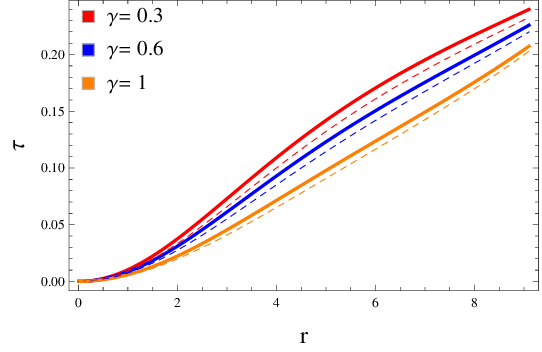,width=0.4\linewidth}
\epsfig{file=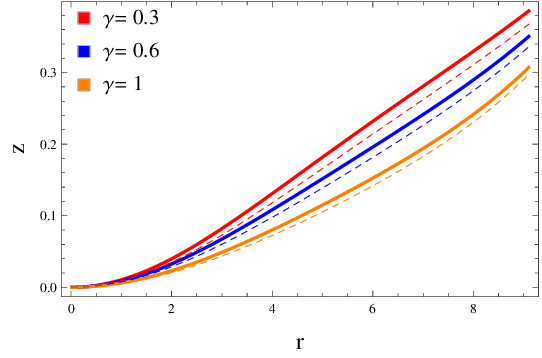,width=0.4\linewidth} \caption{Mass (in
km), compactness and redshift versus $r$ (in km) for
$\bar{\mathcal{S}}=0.1$ (solid lines) and $\bar{\mathcal{S}}=0.8$
(dashed lines) corresponding to $\bar{\Pi}=0$.}
\end{figure}
\begin{figure}\center
\epsfig{file=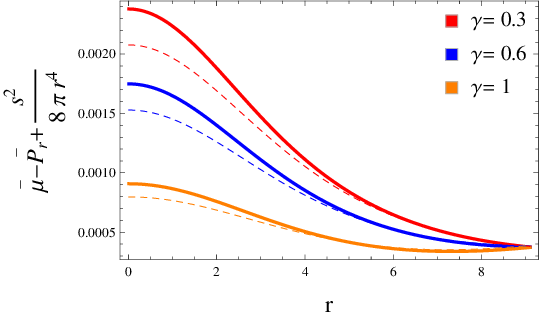,width=0.4\linewidth}\epsfig{file=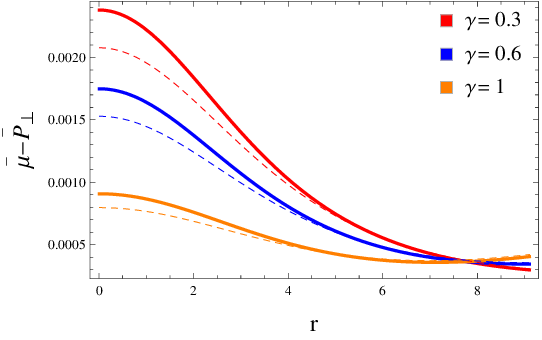,width=0.4\linewidth}
\caption{Dominant energy conditions (in km$^{-2}$) versus $r$ (in
km) for $\bar{\mathcal{S}}=0.1$ (solid lines) and
$\bar{\mathcal{S}}=0.8$ (dashed lines) corresponding to
$\bar{\Pi}=0$.}
\end{figure}
\begin{figure}\center
\epsfig{file=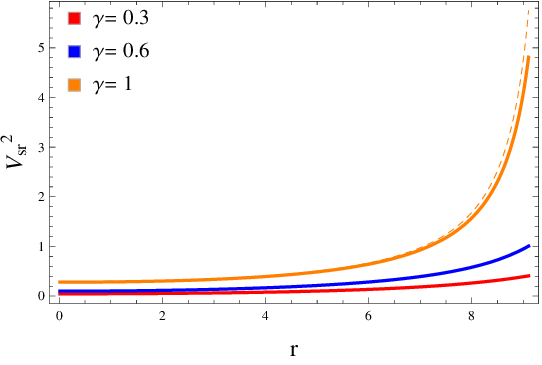,width=0.4\linewidth}\epsfig{file=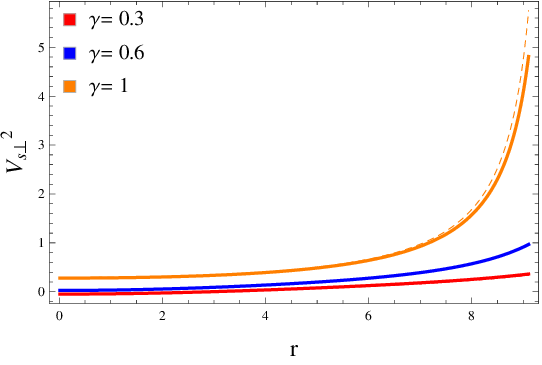,width=0.4\linewidth}
\epsfig{file=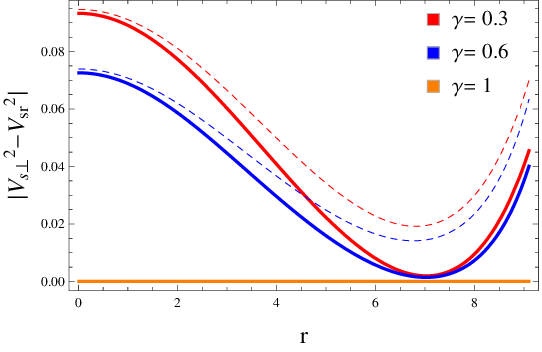,width=0.4\linewidth}
\caption{Radial/tangential sound speeds and cracking criteria versus
$r$ (in km) for $\bar{\mathcal{S}}=0.1$ (solid lines) and
$\bar{\mathcal{S}}=0.8$ (dashed lines) corresponding to
$\bar{\Pi}=0$.}
\end{figure}

Figure \textbf{3} presents the mass function of the corresponding
spherical self-gravitating model. We find the charged/uncharged
isotropic system to be less dense in comparison with the respective
anisotropic analog. Moreover, we calculate the resulting mass at the
spherical boundary in terms of mass of the sun for all parametric
choices and deduced that this model fits best with the observational
mass for $\gamma=0.3$ and $\bar{\mathcal{S}}=0.1$, i.e.,
$\bar{\mathcal{M}}=1.56M_{\bigodot}$. The other two parameters
plotted in the same Figure guarantee the fulfillment of their
respective criteria. The positive trend shown by the state variables
(in Figure \textbf{2}) allows us to plot only the dominant energy
conditions such as $\bar{\mu}-\bar{P}_{r}+\frac{\mathrm{s}^2}{4\pi
r^4} \geq 0$ and $\bar{\mu}-\bar{P}_{\bot} \geq 0$. These bounds can
be seen in Figure \textbf{4} from which the viability of the
resulting model is revealed. Moreover, multiple strategies are used
in Figure \textbf{5} to check whether our solution is stable or
unstable. The upper two plots provide that $\gamma=0.6$ (with
$\bar{\mathcal{S}}=0.1$ and $0.8$) is the only parametric value for
which both sound speeds show stable behavior. Contrariwise, we get
the stable model for every choice of $\gamma$ through the cracking
criteria (lower plot). Figure \textbf{6} exhibits that the modified
scalar factors \eqref{g56a} and \eqref{g60d} decrease and increase,
respectively, by increasing the decoupling parameter. However, both
factors are in an inverse relation with an electric charge. Further,
we compare the behavior of this factor with that of $\mathbb{GR}$
and deduce that the complexity reduces in the former case.
\begin{figure}\center
\epsfig{file=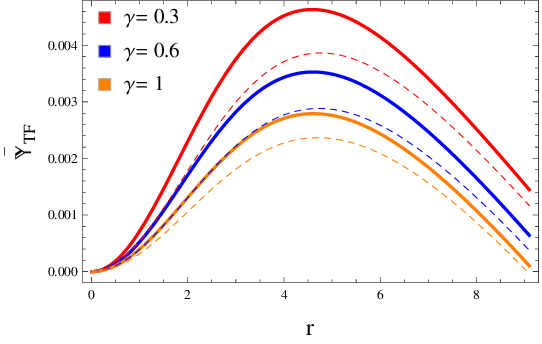,width=0.4\linewidth}\epsfig{file=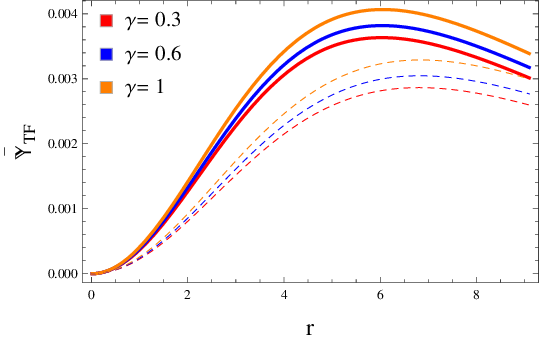,width=0.4\linewidth}
\caption{Complexity factors \eqref{g56a} and \eqref{g60d} (in
km$^{-2}$) versus $r$ (in km) for $\bar{\mathcal{S}}=0.1$ (solid
lines) and $\bar{\mathcal{S}}=0.8$ (dashed lines).}
\end{figure}

The physical properties of the second solution corresponding to
$\bar{\mathbb{Y}}_{TF}=0$ is now explored for the same values of
$\gamma$ and $\chi_3$ along with $\mathbb{C}_{3}=-0.001$. Figure
\textbf{7} shows that the deformed $g_{rr}$ component possesses
singularity-free and increasing nature. The plots of the matter
triplet (provided in Eqs.\eqref{60g}-\eqref{60i}) and anisotropic
factor are demonstrated in Figure \textbf{8}. They own the same
trend as already observed for the first solution. The last plot
exhibits that the core (boundary) of a compact star contains no
(maximum) anisotropy for all chosen parametric values. However, the
increment in charge results in less anisotropy. Figure \textbf{9}
indicates that the greater the value of $\gamma$ and charge, the
less the mass of a compact star is. Further, the required criteria
for the redshift and compactness parameters is fulfilled, as can be
seen from the same Figure. The plots in Figure \textbf{10} guarantee
that the corresponding solution is viable. Figure \textbf{11}
reveals that our resulting model \eqref{60g}-\eqref{60i} is stable
for all parametric choices except $\gamma=1$ (upper left plot).
\begin{figure}\center
\epsfig{file=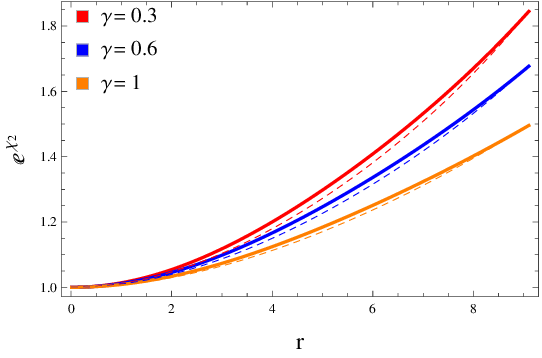,width=0.4\linewidth}
\caption{Deformed $g_{rr}$ metric function \eqref{g60fa} versus $r$
(in km) for $\bar{\mathcal{S}}=0.1$ (solid lines) and
$\bar{\mathcal{S}}=0.8$ (dashed lines) corresponding to
$\bar{\mathbb{Y}}_{TF}=0$.}
\end{figure}
\begin{figure}\center
\epsfig{file=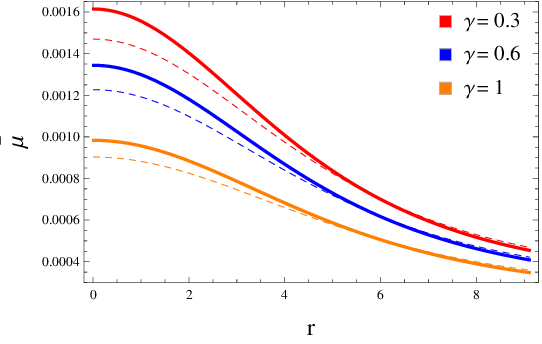,width=0.4\linewidth}\epsfig{file=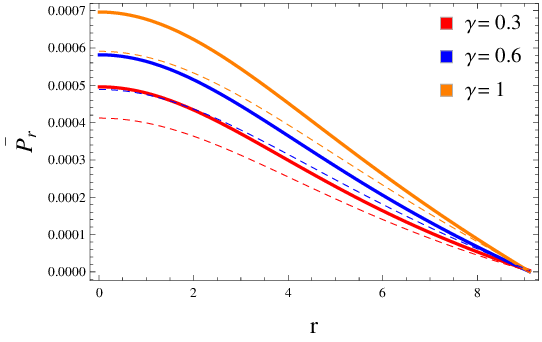,width=0.4\linewidth}
\epsfig{file=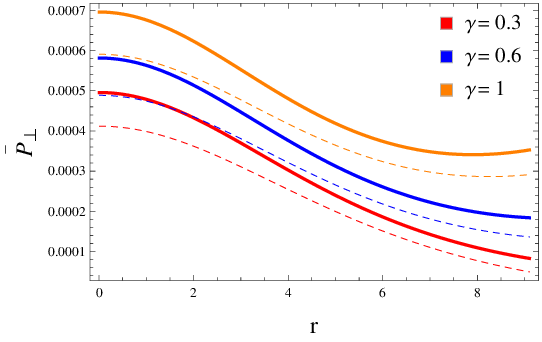,width=0.4\linewidth}\epsfig{file=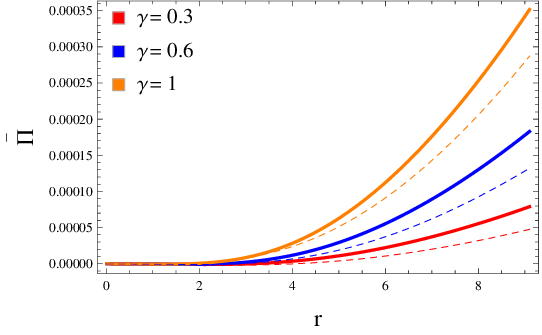,width=0.4\linewidth}
\caption{Matter determinants and anisotropy (in km$^{-2}$) versus
$r$ (in km) for $\bar{\mathcal{S}}=0.1$ (solid lines) and
$\bar{\mathcal{S}}=0.8$ (dashed lines) corresponding to
$\bar{\mathbb{Y}}_{TF}=0$.}
\end{figure}
\begin{figure}\center
\epsfig{file=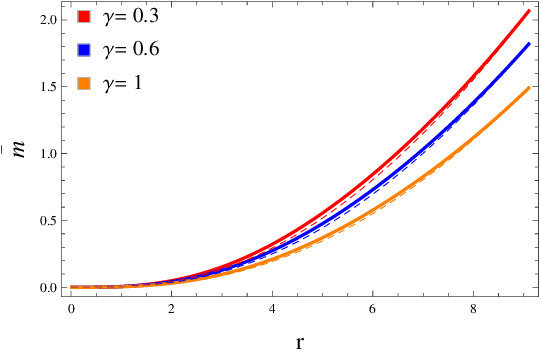,width=0.4\linewidth}\epsfig{file=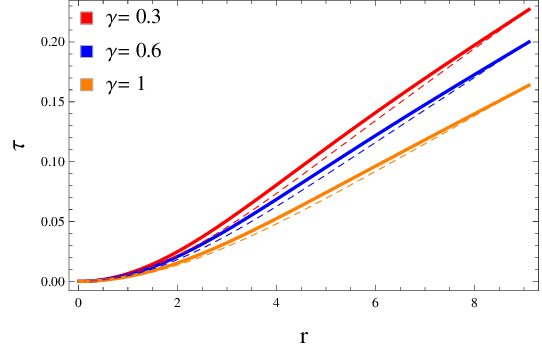,width=0.4\linewidth}
\epsfig{file=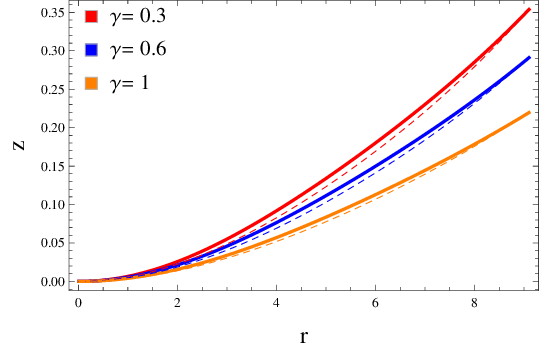,width=0.4\linewidth} \caption{Mass (in
km), compactness and redshift versus $r$ (in km) for
$\bar{\mathcal{S}}=0.1$ (solid lines) and $\bar{\mathcal{S}}=0.8$
(dashed lines) corresponding to $\bar{\mathbb{Y}}_{TF}=0$.}
\end{figure}
\begin{figure}\center
\epsfig{file=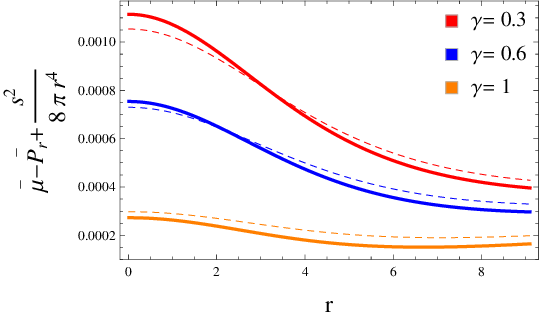,width=0.4\linewidth}\epsfig{file=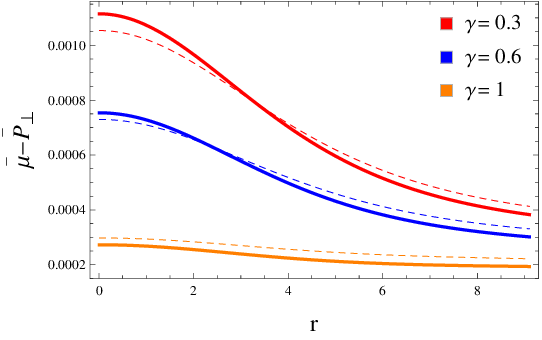,width=0.4\linewidth}
\caption{Dominant energy conditions (in km$^{-2}$) versus $r$ (in
km) for $\bar{\mathcal{S}}=0.1$ (solid lines) and
$\bar{\mathcal{S}}=0.8$ (dashed lines) corresponding to
$\bar{\mathbb{Y}}_{TF}=0$.}
\end{figure}
\begin{figure}\center
\epsfig{file=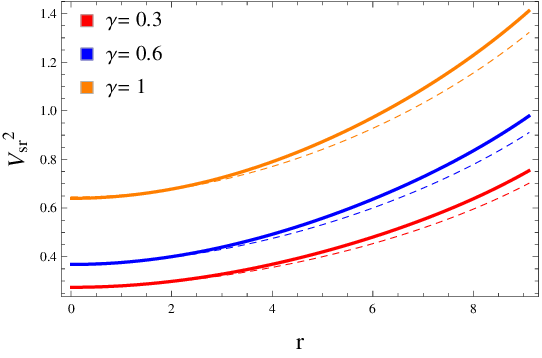,width=0.4\linewidth}\epsfig{file=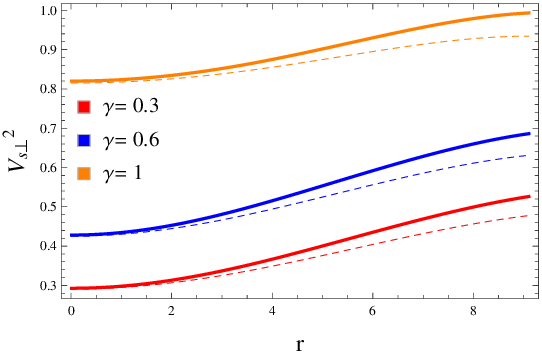,width=0.4\linewidth}
\epsfig{file=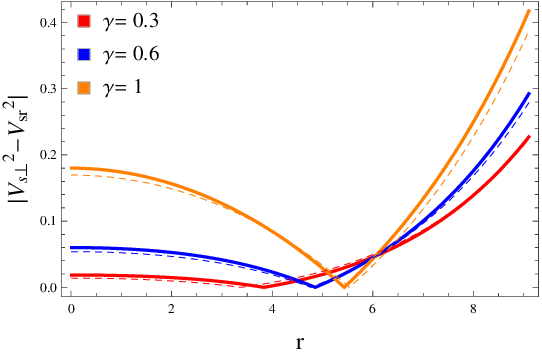,width=0.4\linewidth}
\caption{Radial/tangential sound speeds and cracking criteria versus
$r$ (in km) for $\bar{\mathcal{S}}=0.1$ (solid lines) and
$\bar{\mathcal{S}}=0.8$ (dashed lines) corresponding to
$\bar{\mathbb{Y}}_{TF}=0$.}
\end{figure}

\section{Conclusions}

This paper explores the existence of self-gravitating spherical
anisotropic charged models by extending the original solutions to an
additional matter field by using the strategy of gravitational
decoupling in
$f(\mathbb{R},\mathbb{T})=\mathbb{R}+2\chi_3\mathbb{T}$ theory. We
initially added the Lagrangian densities of the electromagnetic
field and extra source in the action and formulated the
corresponding field equations. We then separated the effects of the
seed and extra matter sources by dividing the field equations into
two individual sets by employing the MGD scheme. The influence of
the charge has only been observed in the field equations
representing the original matter distribution. Two different
solutions to this set were obtained by taking the following ansatz
$$\chi_1(r)=\ln\bigg\{\mathcal{C}_2^2\bigg(1+\frac{r^2}{\mathcal{C}_1^2}\bigg)\bigg\},\quad
\chi_5(r)=e^{-\chi_2(r)}=\frac{\mathcal{C}_1^2+r^2}{\mathcal{C}_1^2+3r^2},$$
and Tolman IV spacetime, respectively. These metric potentials
contain a triplet ($\mathcal{C}_1,\mathcal{C}_2,\mathcal{C}_3$) as
unknown quantities whose values have been calculated through
continuity of the metric potentials at the spherical boundary. The
other set \eqref{g21}-\eqref{g23} encoding the effect of the
additional source was then solved by employing some constraints on
$\mathfrak{C}_{\beta\alpha}$. We have constructed an isotropic model
from being anisotropic for a particular value, i.e., $\gamma=1$,
leading to the first solution. Further, the second model was
developed by taking into consideration that the resultant complexity
factor corresponding to both matter sectors becomes null.

Different values of the decoupling parameter along with fixed values
of the coupling as well as integration constants have been chosen to
graphically analyze the developed solutions so that we can check how
$\gamma$ influences these extended models. The matter determinants
for both solutions \big(presented in \eqref{g46}-\eqref{g48} and
\eqref{60g}-\eqref{60i}\big), their corresponding anisotropic
factors, redshift, compactness and the energy conditions have shown
an acceptable behavior. It was also observed that the first
resulting model corresponding to $\bar{\Pi}=0$ possesses a dense
interior for every value of $\gamma$ in comparison with the second
solution. Further, the radial metric components \eqref{g40a} and
\eqref{g60fa} have been deformed and plotted which exhibit positive
and increasing behavior throughout (Figures \textbf{1} and
\textbf{7}). We have employed the causality and cracking criteria to
analyze whether the modified models show stable behavior for all
choices of the decoupling parameter and charge or not. We have
observed that the first model is stable only for $\gamma=0.6$
through the causality conditions, and thus compatible with the
uncharged $f(\mathbb{R},\mathbb{T})$ framework \cite{25aa} as well
as Brans-Dicke theory only for this value \cite{37m}. However, the
cracking criterion provides both stable solutions. Eventually, all
these results can be reduced to $\mathbb{GR}$ for $\chi_3=0$.\\\\
\textbf{Data Availability Statement:} No Data associated
in the manuscript.\\\\
\textbf{Conflict of Interest Statement:} The authors declare that
there is no conflict of interest.

\end{document}